\documentclass[sigconf]{acmart}
\AtBeginDocument{%
  }

\copyrightyear{2026}
\acmYear{2026}
\setcopyright{cc}
\setcctype{by}
\acmConference[CHI '26]{Proceedings of the 2026 CHI Conference on Human Factors in Computing Systems}{April 13--17, 2026}{Barcelona, Spain}
\acmBooktitle{Proceedings of the 2026 CHI Conference on Human Factors in Computing Systems (CHI '26), April 13--17, 2026, Barcelona, Spain}
\acmPrice{}
\acmDOI{10.1145/3772318.3790716}
\acmISBN{979-8-4007-2278-3/2026/04}

\usepackage{graphicx}
\usepackage{multirow}
\usepackage{subcaption}
\usepackage{cleveref}
\usepackage{array}
\usepackage{enumitem}
\usepackage{float}
\usepackage{makecell}

\begin{document}

\title{The Impact of Response Latency and Task Type on Human-LLM Interaction and Perception}


\author{Felicia Fang-Yi Tan}
\email{felicia.tan@nyu.edu}
\orcid{0000-0001-7837-174X}
\affiliation{%
  \institution{Tandon School of Engineering}
  \institution{New York University}
  \city{New York}
  \country{USA}
}

\author{Moritz A. Messerschmidt}
\email{moritz@ahlab.org}
\orcid{0000-0003-1527-4613}
\affiliation{%
  \institution{Augmented Human Lab}
  \institution{National University of Singapore}
  \city{Singapore}
  \country{Singapore}
}

\author{Wen Yin}
\email{wy870@nyu.edu}
\orcid{0009-0006-9122-9215}
\affiliation{%
  \institution{Tandon School of Engineering}
  \institution{New York University}
  \city{New York}
  \country{USA}
}

\author{Oded Nov}
\email{onov@nyu.edu}
\orcid{0000-0001-6410-2995}
\affiliation{%
  \institution{Tandon School of Engineering}
  \institution{New York University}
  \city{New York}
  \country{USA}
}


\renewcommand{\shortauthors}{Tan et al.}

\begin{abstract}
Responsiveness in large language model (LLM) applications is widely assumed to be critical, yet the impact of latency on user behavior and perception of output quality has not been systematically explored. We report a controlled experiment varying time-to-first-token latency (2, 9, 20 seconds) across two taxonomy-driven knowledge task types (Creation and Advice). Log analyses reveal that user interaction behaviors were robust to latency, yet varied by task type: Creation tasks elicited more frequent prompting than Advice tasks. In contrast, participants who experienced 2-second latencies rated the LLM’s outputs less thoughtful and useful than those who experienced 9- or 20-second latencies. Participants attributed delays to AI deliberation, though long waits occasionally shifted this interpretation toward frustration or concerns about reliability. Overall, this work demonstrates that latency is not simply a cost to reduce but a tunable design variable with ethical implications. We offer design strategies for enhancing human-LLM interaction.

\end{abstract}






\maketitle

\section{Introduction}

Large language model (LLM) assistants like ChatGPT have rapidly become part of everyday knowledge work, commonly supporting tasks such as drafting an email or getting feedback on ideas \cite{pu2025ideasynth,mandal2025utilization, zhang2025friction}. A central factor shaping their usability is responsiveness, long examined in HCI research through the lens of System Response Times (SRT). Classic studies identified thresholds at which response delays disrupt interaction - about 0.1 seconds for immediacy, 1 second for continuity of thought, and 10 seconds for loss of attention~\cite{card_psychology_1983, miller_response_1968}. From this perspective, optimizing for speed and minimizing any waiting that users experience emerged as a core design imperative. Yet in today’s LLM platforms, responsiveness is no longer a simple matter of faster being better. For instance, users are often presented with trade-offs between models that deliver quicker responses vs. slower modes engaging in extended reasoning sequences for greater accuracy~\cite{sui_stop_2025}. In contemporary use cases of LLMs for knowledge work, classical assumptions about latency require reexamination.

Unlike the deterministic, transactional systems typically studied in SRT research, waiting in LLM interactions introduces new dynamics. Because outputs are probabilistic and unpredictable, users cannot know in advance what the model will generate or how well it will align with their goals, a difficulty Subramonyam et al.\ term the “gulf of envisioning”~\cite{subramonyam_bridging_2024}. Moreover, interaction with LLMs takes conversational form, which suggests that timing may carry socially meaningful cues, much as pauses shape interpretations in human dialogue~\cite{hwang_when_2019,lee_while_2025}. Prior work points to the possibility that delays can be anthropomorphized as “thinking time” or even serve constructive functions when calibrated as positive friction~\cite{chen_exploring_2024,inan_better_2025,stackpole_help_2024}. What remains less understood, however, is how such differences in interpretation play out in everyday knowledge work, where users pursue different task types and may respond to waiting differently.

This motivates our focus on task type as a critical moderator of how latency is experienced and shapes user behaviors in LLM-assisted knowledge work. We focus on two common forms of knowledge work~\cite{brachman_how_2024}: \emph{Creation} tasks, which involve producing new content (e.g., writing, brainstorming, or drafting), and \emph{Advice} tasks, which involve providing guidance, critique, or evaluation (e.g., suggesting revisions, offering decision support, or validating criteria). Based on this distinction, our study was guided by two research questions:

\begin{enumerate}[label= \textbf{RQ\arabic*.}, topsep=8pt]
    \item How do Latency and Task Type influence human-LLM interaction behavior?
    \item How do Latency and Task Type affect participants’ perceived quality of LLM responses?
\end{enumerate}

To investigate these questions, we conducted a 3 (Latency: 2, 9, 20 seconds) × 2 (Task Type: Creation, Advice) between-subjects experiment with 240 participants. Using a custom GPT-4o interface, participants were assigned to one latency condition and one task type, then completed three tasks within their assigned task type. Each time they queried the AI assistant, the response was delayed according to their assigned latency condition. We examined system-logged behaviors (e.g., prompt submissions and edits, copy-pasting) alongside user perceptions gathered through ratings and open-ended reflections.

Results show that participants' log-based interaction behaviors were robust to latency but strongly shaped by task type. Across conditions, participants in Creation tasks submitted more new prompts than those in Advice tasks, underscoring that the kind of work users pursue with LLMs drives their prompting strategies more than system response times. Latency, in contrast, primarily influenced users’ perceptions of output quality: participants who experienced short waits (2\,s) rated the LLM’s responses as less thoughtful and less useful than those generated after longer waits (9-20 s). Qualitative feedback echoed this pattern, with participants often interpreting longer response latencies as signs of ``deliberation'' rather than strain. Taken together, these findings nuance the maxim that faster is better.

This work makes three key contributions:

\begin{enumerate}[nosep]
    \item We present empirical findings showing that interaction behaviors were robust to latency but shaped by task type. In contrast, both latency and task type influenced perceptions of LLM output quality, with short waits (2\,s) rated as less thoughtful and useful than longer waits (9-20\,s).
    \item We provide design implications for treating timing as an interaction design material, suggesting how response latency can be framed and tuned based on task structure.
    \item We contribute a reusable experimental platform for studying timing effects in human-LLM interaction, comprising an end-to-end LLM interaction system with latency control and interaction logging\footnote{https://github.com/AugmentMo/gpt-chatbot-emu}, and the study’s anonymized dataset with accompanying documentation on the Open Science Framework (OSF)\footnote{https://osf.io/367cw}.

\end{enumerate}

\section{Related Work}

\subsection{System Response Times in HCI}
\label{RelatedWork-System Response Times in HCI}
    Early HCI studies often frame system delays in conversational and cognitive terms. Miller’s analysis of dialogue suggests that multi-second pauses ($\approx$4 s) can break the interaction ``thread'' \cite{miller_response_1968}, i.e., disrupt users' train of thought and mental flow. Usability guidance later converged on thresholds of immediacy: about 0.1 s to feel instantaneous (preserving the illusion of direct manipulation), $\sim$1 s to maintain the flow of thought, and $\sim$10 s before attention drifts \cite{nielsen_usability_1994, card_psychology_1983}. These anchors, spanning perceptual, cognitive, and attentional scales, suggest a dominant heuristic that the faster a system responds, the more seamless the user experience.
    
    Subsequent research showed that latency affects not only task completion speed but also error rates, user strategies, satisfaction, and physiological stress markers. Dabrowski and Munson \cite{dabrowski_40_2011} classify tasks by their tolerance for delay. Control tasks, such as dragging, scrolling, or other direct-manipulation actions, require near-instantaneous feedback ($<$200 ms) to preserve the feeling of continuous control. In contrast, conversational tasks admit modest pauses; short latencies can feel like natural turn-taking, sustaining rhythm rather than disrupting it. A controlled study of interactive visualization showed that adding 500 ms of delay to each action reduced coverage (how broadly people explored the data) and insight (the number or quality of observations they made) \cite{liu_effects_2014}.
    
    Two additional factors refine this picture: the psychophysiology of waiting and the role of expectations. Human-factors experiments indicate a U-shaped cost curve, where short $\sim$2\,s waits can elevate arousal and impair accuracy, whereas 8-9 s waits increase strain and frustration \cite{thum_standardized_1995, kohlisch_system_1997}. Expectations amplify these effects. In high-volume services such as web search, where users are habituated to near-instant responses, delays above roughly 1 s are noticed, and users preferentially click on the faster-loading page even when the content is identical \cite{arapakis_impact_2014}.
    
    Together, these studies yield a long-standing baseline: treat latency as overhead to minimize. However, these thresholds are not fixed but moderated by task class, user expectations, and whether the system provides informative waiting cues.

\subsection{Latency Perception, Mental Models, and Social Heuristics}
\label{RelatedWork-Latency Perception, Mental Models..}

    Unlike deterministic systems where an action yields a predefined outcome (e.g., opening a file), LLMs generate probabilistic outputs, respond in conversational turns. In this setting, time is a communicative signal of effort, or malfunction that shapes how people trust, strategize, and collaborate with intelligent systems \cite{hwang_when_2019, kruger_effort_2004, riedl_system_2018, logg_algorithm_2019}. Brief pauses are often interpreted as ``system working'' cues, whereas sustained waits are read as deliberation when complexity is expected, or as failure when speed is normative \cite{nielsen_usability_1994}.
    
    Classic cognitive accounts hold that people construct simplified internal models to explain complex systems \cite{norman_observations_1983}. Such models guide users’ inferences about what the system is doing while they wait, often becoming folk theories in opaque algorithmic settings. These mental models also shape pacing: fast responses elicit fragmented, exploratory probing, whereas slower responses encourage fewer, more deliberate commands \cite{dabrowski_40_2011}. Managing latency is thus also about managing perception - what time signals about effort, control, and competence shapes how interaction unfolds.
    
    Furthermore, classic social response theory explains that people apply conversational norms to machines \cite{nass_machines_2000}. Recent studies report similar attributions in AI-mediated dialogue; Lee et al. \cite{lee_while_2025} found that users of creative image-generation tools often interpreted longer render times as evidence that the AI was ``thinking harder,'' sometimes rating outputs as higher quality despite identical content. Subramonyam et al. \cite{subramonyam_bridging_2024} highlight a related ``gulf of envisioning'' in LLM use. Because users lack direct visibility into model processes, they imbue observable cues such as delays with explanatory weight. Crucially, whether latency is judged positively or negatively depends on context and framing. Delays in speed-critical settings (e.g., factual Q\&A) may be seen as incompetence or system failure. But when delays align with expectations of complexity, they can be reassuring. Zhang et al.\ \cite{zhang_explaining_2024} demonstrate that simple framing cues (e.g., displaying ``Analyzing your request...'' during a pause) increased perceived transparency and trust. Riemer et al.\ \cite{riemer_time_2023} add that temporal consistency matters. In other words, once users develop expectations from repeated encounters, deviations (i.e., unexpectedly fast or slow responses) can erode confidence. 

      While the aforementioned studies clarify how users perceive and interpret latency, the meaning and impact of waiting are also contingent on what users are trying to accomplish.

\subsection{Task Contexts in LLM Use}
\label{RelatedWork-Task Contexts in LLM Use}

   LLMs support a spectrum of knowledge-work tasks that differ in cognitive demands and evaluation criteria \cite{mendel2025laypeople}. Prior work organizes task context along three empirically observed axes: stakes, expectations, and interaction patterns. Stakes vary with domain and the consequence of error. Unreliability is especially concerning in high-stakes settings compared with low-risk or exploratory uses \cite{brynjolfsson_generative_2025}. Expectations vary by intent: advice-oriented tasks prioritize factuality and professional knowledge, favoring concise, actionable output, whereas creation-oriented tasks prioritize creativity and favor more expansive, stylistically flexible output \cite{wang_task_2024, brachman_current_2025}. Interaction patterns also differ. Advice and learning intents tend to unfold through multi-turn conversational prompting, which supports clarification and iteration, whereas information-seeking often takes the form of single-turn lookups or tool-assisted retrieval \cite{shah_using_2025, gao_taxonomy_2024}. 
    
    Within this landscape, Brachman et al.\ \cite{brachman_current_2025} validate a four-way taxonomy of LLM use -- Creation, Information, Advice, and Automation. Our study focuses on Creation and Advice, two major task types in knowledge work. Creation tasks include writing an email, brainstorming ideas, or drafting code or prose, while Advice tasks involve offering critiques, making recommendations, or helping a user weigh trade-offs. These categories differ in cognitive orientation, with Creation emphasizing open-ended generativity and originality, and Advice emphasizing evaluative guidance and reliability.
    
    Empirical contrasts reinforce that latency may be interpreted differently by task type. Creation emphasizes novelty and fluency through iterative prompting, whereas Advice emphasizes precision and verification \cite{wang_task_2024, shah_using_2025}. A 10 s pause may feel reasonable when “the AI is coming up with ideas,” but excessive when the goal is a binary recommendation. Despite this plausible interaction, latency research rarely accounts for task diversity. Controlled studies often target narrow domains or single tasks, while taxonomy-driven work maps LLM use without investigating how timing operates across categories. This leaves a critical gap in understanding latency effects across task types.

\subsection{Positive Friction and Productive Delay}
\label{RelatedWork-Positive Friction and Productive Delay}

    \emph{Positive friction}, an HCI design framework, reframes small, intentional slowdowns as levers that interrupt automatic responding and encourage more deliberate, effortful reasoning (often described as System~2 processing \cite{stanovich_individual_2000, kahneman_thinking_2011}). Chen and Schmidt \cite{chen_exploring_2024} distinguish four friction types (self-control, questioning assumptions, stimulating action, and deprioritizing efficiency), each implementable via forcing functions, pauses, or prompts. In time-based applications, well-framed delays can improve decision quality, yielding better discrimination of correct vs.\ incorrect advice and reducing overreliance on AI \cite{park_slow_2019, bucinca_trust_2021}.
    
    Concrete implementations of positive friction appear in human-AI interaction. Inan et al.\ \cite{inan_better_2025} focus on conversational AI, introducing brief pauses and clarifying interjections in dialogue to improve alignment. A study also showed that color-coded highlights indicating confidence levels, omissions, or contested claims, function as cognitive ``speed bumps,'' slowing readers enough to encourage critical review \cite{stackpole_help_2024}. Notably, effects followed an inverted-U: too little friction left users skimming, whereas too much imposed unnecessary burden.
    
    Foundational timing work also explains how delay paces cognition via (1) \emph{entrainment}, where users synchronize their tempo to system speed, so very short waits rush and increase errors while moderate waits support control \cite{myer_towards_2002}, and (2) \emph{temporal predictability}, where consistent waiting times allow users to prepare actions in advance, improving performance \cite{thomaschke_predictivity_2014}.

\vspace{2mm}

In summary, prior work underscores four themes: (1) classic HCI established low latency as a core usability principle, (2) latency can be interpreted through users’ mental models and social heuristics, (3) the meaning and acceptability of delay depend on task context, and (4) deliberate slowdowns have been framed as positive friction. Our work extends this literature by systematically comparing latency across task types in LLM-assisted knowledge work.

    \begin{figure*}[htbp]
      \centering
      \includegraphics[width=0.9\textwidth, trim=5 5 5 5, clip]{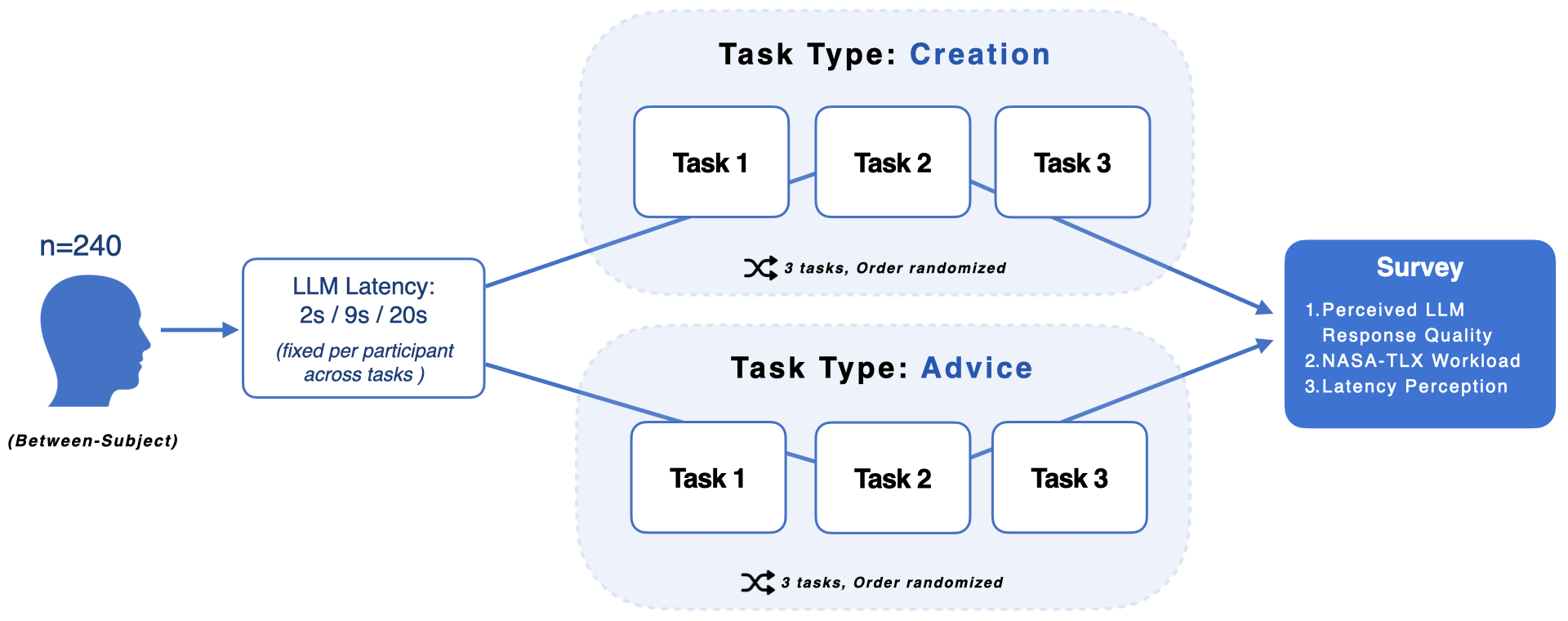}
      \caption{Study design. Participants were randomly assigned to one of six experimental groups (2 Task-Types $\times$ 3 Latency-Levels). Each participant completed three tasks of the assigned Task-Type, in randomized order, interacting with the LLM under the assigned latency, which remained fixed across all three tasks. System logs recorded user interactions while post-task surveys captured self-reports.}
        \label{fig:experimentaldesign}
        
        \Description{Flow diagram of the study showing random assignment to latency and task type, three randomized tasks, then a survey. Starting with 240 participants, each is randomly assigned one latency level (2, 9, or 20 seconds; between-subjects) and one task type (Creation or Advice; between-subjects). Within the chosen task type, the participant completes three LLM tasks at the assigned latency, with task order randomized. The diagram ends with a survey that measures perceived response quality of LLM responses, NASA-TLX workload, and perceived LLM delay. Visual grouping highlights the Creation and Advice task blocks; each task box is labeled “LLM (assigned latency),” and a shuffle icon denotes randomized order.}
    \end{figure*}

\section{Study}

We conducted a controlled experiment to examine how task type and latency shape human–LLM interaction and perception. This section presents the study design and details the system implementation used for latency control and interaction logging.

\subsection{Study Design}

The experiment followed a 2 $\times$ 3 between-subjects design with 240 crowdsourced participants. Participants were randomly assigned to one of the following factors (Figure~\ref{fig:experimentaldesign}):
        \begin{itemize}
            \item \textbf{Task-Type:} Creation, Advice
            \item \textbf{Latency-Level:} 2, 9, 20 seconds
        \end{itemize}

\subsubsection{Task-Type}

     Participants were randomly assigned to one of two task types (Creation or Advice) drawn from Brachman et al.’s taxonomy \cite{brachman_current_2025}. These two categories were chosen for their prevalence in knowledge work and distinct cognitive orientations (Section~\ref{RelatedWork-Task Contexts in LLM Use}). \textit{Creation} tasks involve producing new content or artifacts (e.g., writing an email, or brainstorming ideas). They are generative, often open-ended, and emphasize originality. By contrast, \textit{Advice} tasks involve evaluative or assistive guidance (e.g., critiquing text, or making a recommendation). They foreground accuracy and reliability, often carry higher stakes of “getting it right,” and typically unfold as dialogic, back-and-forth exchanges. Together, these task types differ in cognitive demands, urgency, and temporal expectations, making them a useful contrast.

    \paragraph{{Task Scenarios}}
    
       Within their assigned Task-Type, participants completed three task scenarios (i.e., three Creation or three Advice scenarios). Task scenarios served as concrete instantiations of Task-Types. For each task scenario, participants freely interacted with the LLM for assistance in completing the task, then submitted a final written response to the task.
    
        The scenarios were adapted from prior HCI/AI task designs \cite{wang_task_2024,shah_using_2025}. Using multiple distinct scenarios per Task-Type improved external validity by ensuring conceptual coverage and reducing the likelihood that findings were unique to a single case. Below, we summarize the scenarios; full task texts are provided in Appendix~\ref{appendix-Experimental tasks}.
        \vspace{1mm}
        
        \textbf{Creation tasks:}
        \begin{enumerate}[nosep]
        
            \item Slogan generation: create and refine a promotional slogan for a fictional collaboration platform targeted at educators.
            \item Scenario-based brainstorming: propose activity ideas to help distracted high school students stay on task.
            \item Gap-filling writing: complete the missing middle section of an article on the benefits of learning a second language.
        \end{enumerate}
        
        \textbf{Advice tasks:}
        \begin{enumerate}[nosep]
            \item Personalized recommendation: advise a coworker struggling with focus while working from home.
            \item Trade-off analysis: evaluate a decision memo comparing a promotion vs.\ a new job offer, identifying missing considerations.
            \item Mock email reply: improve a draft response to a friend considering a career change to become a full-time artist.
        \end{enumerate}
    \vspace{3mm}
    Within each Task-Type, the three scenarios were presented in simple random order.

\subsubsection{Latency-Level}

    \paragraph{{Latency Definition}}
    
        We varied time-to-first-token (TTFT), defined as the interval between prompt submission and rendering of the first generated token. We adopt TTFT as an onset-based proxy for responsiveness, consistent with classic HCI definitions of response time as the interval between user input and initial system feedback \cite{miller_response_1968,card_psychology_1983,nielsen_usability_1994}.
        
         This manipulation was implemented via a custom HTML/JavaScript front-end and a Python back-end, which fixed TTFT at one of three levels (2, 9, and 20 seconds). After the first token, the streaming rate was fixed at 25 tokens/second (40 ms inter-token interval) in all conditions. Overall output length followed the model’s natural decoding process, allowing total generation time to vary with response length. Section~\ref{System Design} provides additional details on the technical implementation. This design isolates TTFT as the focal manipulation while holding other temporal components constant. Unmanipulated factors are discussed as directions for future work (Section~\ref{Limitations and Future Work}).

        \begin{figure*}[htbp]
              \centering
              \includegraphics[width=0.9\textwidth]{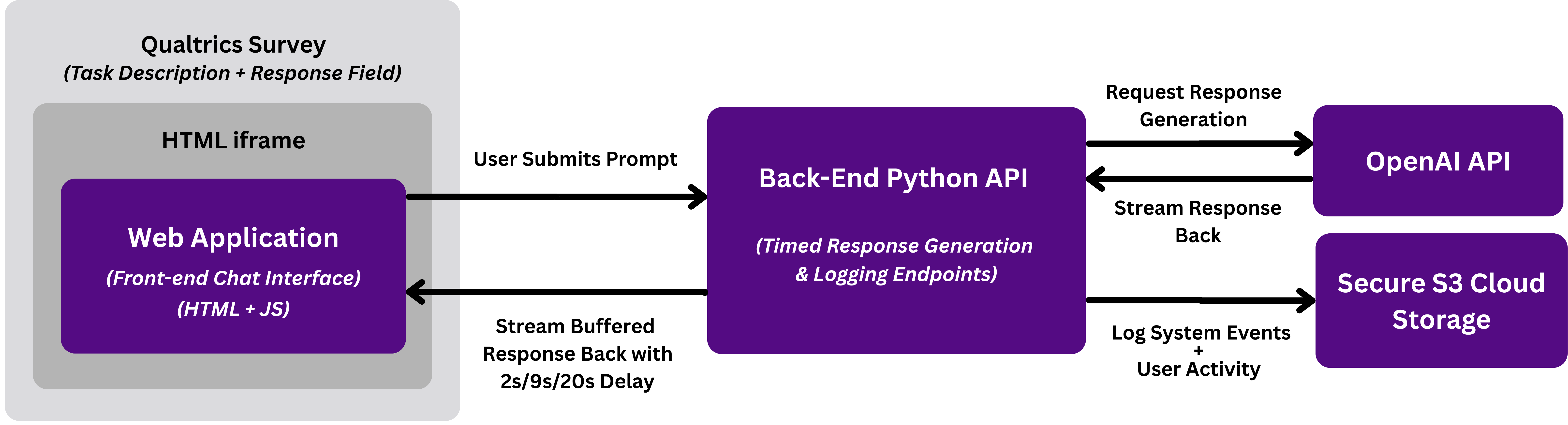}
              \caption{System architecture. The Qualtrics survey embeds a web application via an HTML iframe. The front-end displays a chat interface, while the Python back-end relays model calls to OpenAI. The back-end then buffers the response stream to enforce the assigned TTFT latency and writes anonymized interaction logs to a secure S3 bucket.}
                \label{fig:architecture}
                
                \Description{Block diagram of the embedded LLM assistant. A Qualtrics survey embeds an HTML iframe that loads a web application. Within the web app, a front-end chat interface (HTML + JavaScript) connects to a back-end Python API. The back-end sends model requests to the OpenAI API, buffers the streamed output to enforce assigned time-to-first-token (TTFT) latencies of 2, 9, or 20 seconds, and streams responses back to the front-end. The back-end also writes anonymized system events and user activity logs to secure Amazon S3 storage.}
        \end{figure*}

    \paragraph{Rationale for Latency Levels}
    
    Our latency levels extend thresholds identified in prior HCI research (Section~\ref{RelatedWork-System Response Times in HCI}), adapting them to LLM interaction dynamics. In this literature, heuristic guidelines at the scale of 1 s and 10 s denote approximate timescale regions where qualitative shifts in user experience occur, rather than precise psychophysical cutoffs. We therefore use 2\,s, 9\,s, and 20\,s to instantiate short, mid-range, and longer delay timescales.
    
    \begin{itemize}[nosep]
        \item \textbf{2\,s (low).} Prior work suggests that response delays of $\sim$1--2\,s can preserve conversational flow and continuity of thought \cite{miller_response_1968}. Because current LLMs rarely respond in sub-second timescales, we anchored our shortest latency level at 2\,s, which serves as a practical lower bound.
        
        \item \textbf{9\,s (moderate).} Positioned within the 8--9\,s range where psychophysiological studies report elevated strain and frustration \cite{thum_standardized_1995,kohlisch_system_1997}, yet just below the widely cited $\sim$10\,s guideline for attentional break \cite{nielsen_usability_1994}, this level probes user experience in a range where the character of waiting is expected to shift.

        \item \textbf{20\,s (high).} Chosen as a stress-test condition, this level reflects realistic response times for reasoning-oriented LLMs under long contexts or inference-time scaling \cite{lin_sleep-time_2025}. It exceeds the $\sim$10\,s attentional window and approaches the $\sim$15\,s region where context switching becomes likely \cite{nielsen_usability_1994}, enabling us to examine behavior beyond established continuity limits.
    \end{itemize}

    \subsubsection{Pilot validation} 
    We conducted three pilot studies (pilot 1: $n=20$ participants; pilot 2: $n=30$; pilot 3: $n=30$) to iteratively refine study materials and verify factor levels. Each pilot recruited a unique set of participants, and participants who completed any pilot study were excluded from the main experiment. Across pilots, the 2/9/20\,s delays were tolerable and subjectively distinguishable; tasks and instructions were clear, and the end-to-end procedure ran smoothly. Minor wording and layout adjustments were made. Pilot data were excluded from analysis, as they served only to ensure feasibility of the final experiment.

\subsection{System Design}
\label{System Design}

    \subsubsection{Architecture Overview}

We developed a system (Figure~\ref{fig:architecture}) that embedded an LLM assistant within a Qualtrics survey, enabling natural and intuitive chat-style interaction while allowing us to control time-to-first-token (TTFT) latency and capture detailed logs.

    \subsubsection{Front-End Interface}

        \begin{figure*}[htbp]
            \centering
            \includegraphics[width=0.61\textwidth, trim=1 7 1 3, clip]{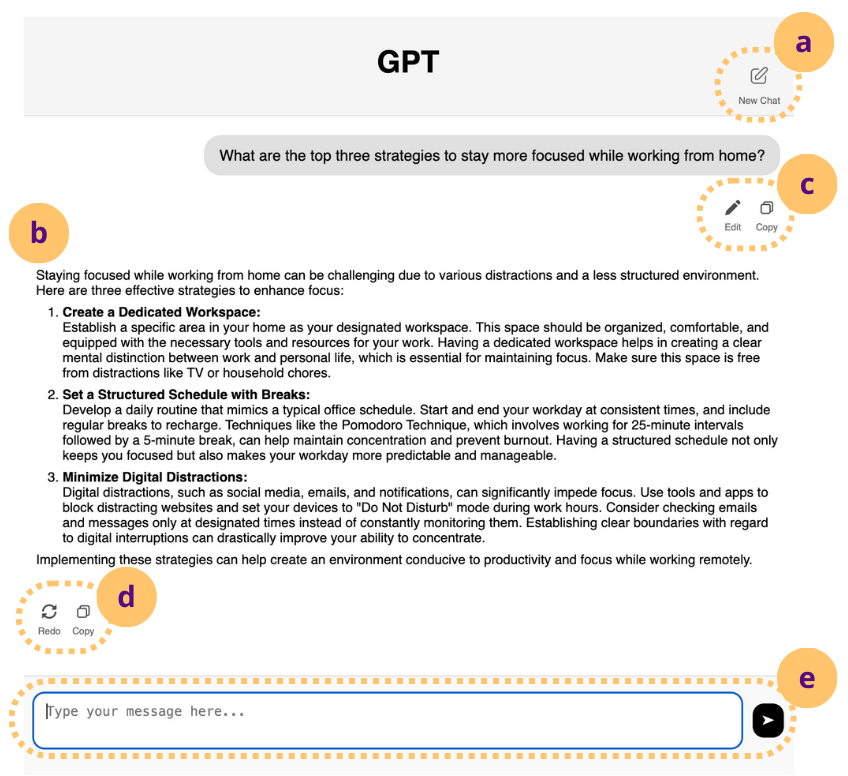}
            \caption{Front-end chat interface with key features highlighted: (a) start or refresh a new chat, (b) chat view with streamed responses, (c) edit or copy messages previously input by the user, (d) regenerate or copy responses generated by the model, and (e) input field to submit new prompts.}
            \label{fig:FrontendChat}
            
            \Description{Mockup of a front-end chat interface with five labeled features. (a) “New Chat” control in the header. (b) Chat view showing a user question, “What are the top three strategies to stay more focused while working from home?”, and a streamed AI reply listing three strategies: create a dedicated workspace, set a structured schedule with breaks, and minimize digital distractions. (c) Inline controls to edit or copy the user’s message. (d) Inline controls to regenerate or copy the model’s response. (e) Prompt input field with a send button at the bottom.}
        \end{figure*}      
    
    The front-end mirrored common LLM assistants (e.g., ChatGPT) and supported six functions:  
    
    \begin{itemize}[nosep]
        \item \textbf{Prompt submission} via a text field and send button (Figure~\ref{fig:FrontendChat}e).  
        \item \textbf{Chat history view}, where responses were streamed token-by-token and rendered using the \emph{marked.js} library for Markdown-to-HTML and \emph{highlight.js} for code syntax highlighting (Figure~\ref{fig:FrontendChat}b).  
        \item \textbf{Edit previous prompt} (Figure~\ref{fig:FrontendChat}c).  
        \item \textbf{Copy message} (Figure~\ref{fig:FrontendChat}c \& d).  
        \item \textbf{Regenerate response} (Figure~\ref{fig:FrontendChat}d).  
        \item \textbf{Start new chat} to clear the conversation (Figure~\ref{fig:FrontendChat}a).  
    \end{itemize}

    During response generation, all interactive controls (submit, edit, regenerate, start-new-chat) were temporarily disabled and visually muted. A pulsing cursor signaled pending output until the assigned TTFT elapsed. The front-end handled all participant inputs and displayed outputs, while forwarding requests to the back-end for latency enforcement.

    \subsubsection{Back-End Python API}
    \label{Backend python}
    
        The Python back-end was hosted on \emph{Render.com} and wrapped the OpenAI API to enforce TTFT latency. When the front-end submitted a request, it passed the current chat history, user message, and assigned TTFT value to the server. The back-end then initiated a streamed GPT-4o response but buffered output in memory until the designated TTFT (2\,s, 9\,s, or 20\,s) had elapsed. After this TTFT delay, tokens were released at a fixed rate of 25 tokens/s (40\,ms inter-token interval). 
        GPT-4o was selected as it was the default model provided in the free tier of ChatGPT at the time of the study, making it familiar to most participants.  
        
        In addition to LLM response generation, the back-end exposed logging endpoints. All model queries and user action events (Table \ref{systemlogevents}), along with timing metadata were recorded. Data were securely stored on AWS S3, retaining only a session identifier.

    \subsubsection{Integrated Participant Interface}
    
             \begin{figure*}[htbp]
              \centering
              \includegraphics[width=0.455\textwidth, trim=0 18 0 18 clip]{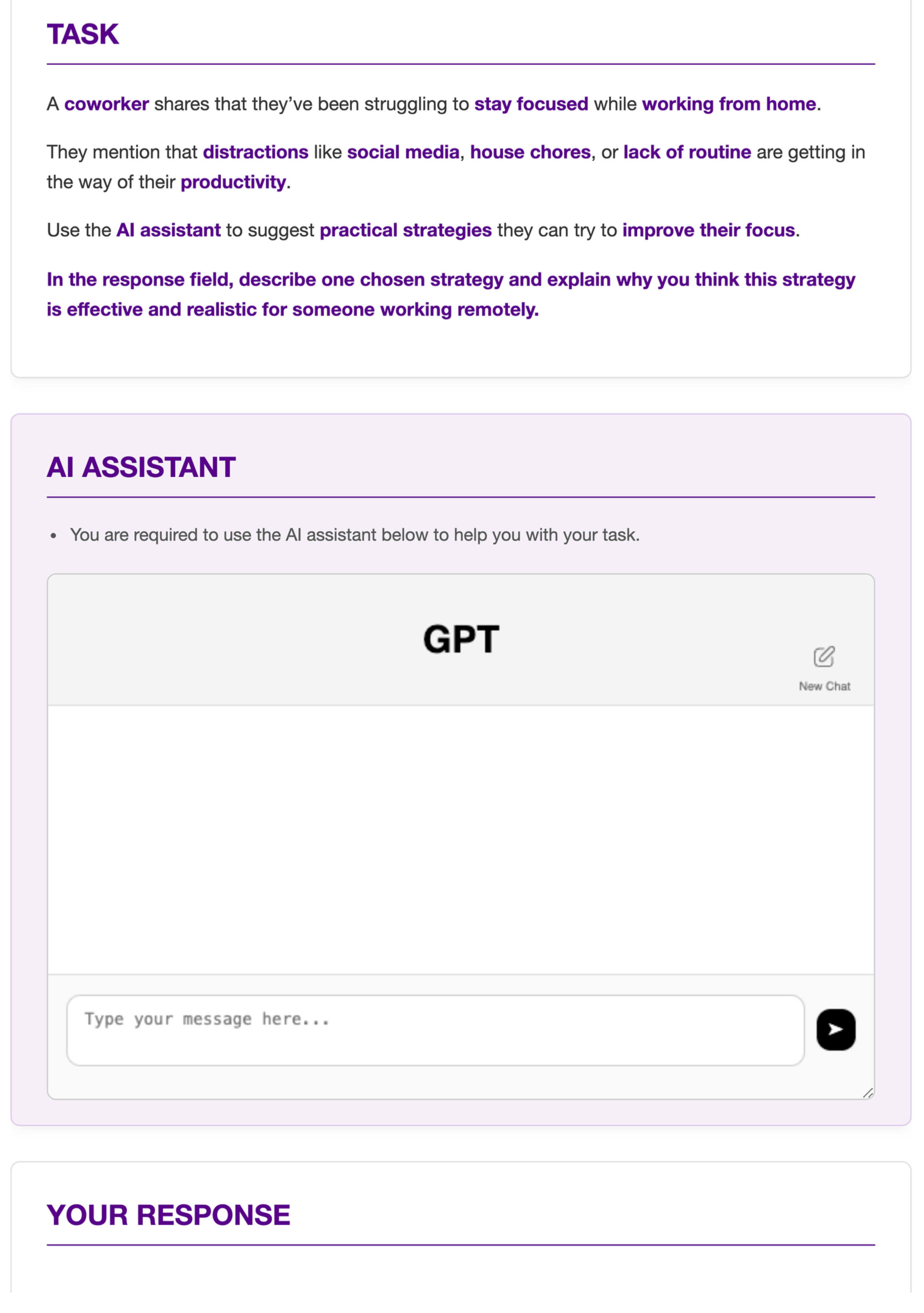}
              \caption{The experiment's task interface. The top panel presents the task description, the middle panel is where participants interact with the AI assistant, and the bottom panel contains the response field where participants submit their final answer.}
                \label{fig:taskinterface}
                
                \Description{Three-panel task interface. The top panel displays the task description and instructions. The middle panel provides the AI assistant chat where the participant interacts with the LLM. The bottom panel contains a text field for entering and submitting the participant’s final answer. The stacked layout presents context, interaction, and submission within a single vertical view.}
            \end{figure*}

    We presented the aforementioned LLM chat module as part of an integrated layout that combined three elements in the overall experimental interface (Figure~\ref{fig:taskinterface}):

    The overall experimental interface comprised:
    \begin{enumerate}[nosep]
        \item \textbf{Top panel:} Task description outlining the scenario and instructions.
        \item \textbf{Middle panel:} LLM chat interface, which functions as the AI assistant for task support.
        \item \textbf{Bottom panel:} A submission box for participants’ final answers.
    \end{enumerate}

\vspace{2mm}
    This arrangement collocated task context, LLM chat assistant interaction, and participant task authoring, minimizing context switching and ensuring that latency manipulations were experienced within a consistent task workflow. 
    
    Text selection and copying from the task description were disabled (Figure~\ref{fig:taskinterface}, top panel). However, copy–paste functionality remained enabled within the chat interface. This prevented participants from transferring the full text description directly into the AI assistant, but allowed for the reuse or adaptation of AI-generated text in the final submissions.

    \subsubsection{Web Application Hosting \& Data Privacy}

    We used Prolific to recruit participants and assign them to experimental conditions, with session tokens and condition data passed to Qualtrics via URL parameters. The web app (front-end + back-end) was hosted on \emph{Render.com} and embedded in Qualtrics via an HTML iframe. Communication between Qualtrics and the app was established through cross-document messaging, allowing session tokens, task type, and latency level to be passed.
    
    All communication occurred over secure HTTPS connections, and only anonymized session tokens linked to our survey responses were logged by our system. OpenAI does not use API data for model training\footnote{https://platform.openai.com/docs/guides/your-data}.

\subsection{Procedure}

    Participants were recruited online via Prolific\footnote{\url{https://www.prolific.com}} and entered the Qualtrics survey. Participants were informed that they had to use a laptop or desktop device, and complete the study in one sitting without refreshing or closing the browser. The study then unfolded in the following stages:
    
    \begin{itemize}
        \item \textbf{Consent and Instructions.} After IRB-approved consent, participants were instructed that they would complete three text-based tasks using an AI assistant. They were introduced to the available chat features (e.g., regenerate, edit, start new chat, copy–paste), informed that queries had to be text-only, and told they could interact with the assistant as many times as needed.  
        
        \item \textbf{Task Block.} Participants completed three tasks within their assigned task type, presented in randomized order.
        
        \item \textbf{Post-task Survey.} After completing all tasks, participants provided ratings and open-ended reflections (Section~\ref{Measures}). 
        
        \item \textbf{Demographics and Debrief.} Participants provided demographic information and were debriefed about the experiment’s purpose, including the option to withdraw their data.  
    \end{itemize}

    The median completion time was 25 min 12 s.

\subsection{Participants}
\label{Participants}

    On Prolific, we restricted eligibility to adults (18+) residing in the United States. Participants self-reported English proficiency, had at least 100 prior submissions with a $\geq$95\% approval rate, and reported using AI assistants (e.g., ChatGPT, Google Gemini) at least once per month for knowledge tasks such as writing or decision-making.  
    
    A total of 240 participants were randomized into six experimental conditions, with group sizes as follows:
    
    \begin{itemize}
        \item Creation: 2 s ($n=41$), 9 s ($n=42$), 20 s ($n=41$)
        \item Advice: 2 s ($n=38$), 9 s ($n=38$), 20 s ($n=40$)
    \end{itemize}
    
We analyzed data from 309 participants with valid interaction logs\footnote{Of 343 participants who finished the survey, 34 lacked a valid log due to an intermittent session-token handoff failure during page load. This server-side failure occurred before any model calls and was not associated with Latency or Task assignment.}. From these, we excluded participants who withdrew consent (n=3), failed the attention check (n=1), or exhibited speeding (n=2). We then applied a latency-fidelity rule excluding participants whose median observed latency exceeded 120\% of their assigned latency ($n=63$; see Section \ref{Measures-InteractionLog} on rationale and preprocessing method). The final analytic sample of 240 met the requirement from our a priori power analysis for a 2$\times$3 between-subjects design (Cohen’s $f=0.25$, $\alpha=.05$, power = .80), which required a sample of $N=211$ ($\sim$36 per condition). Participant demographics for the final analytic sample are reported in Appendix~\ref{participant demographics}.

Participants were compensated \$4 USD for completing the study (median completion time = 25 min 12 s). The study protocol was approved by our institutional review board (IRB).
    
    \vspace{-2mm}

\subsection{Measures}
\label{Measures}

We collected both behavioral and subjective measures to capture how latency shaped interaction strategies (RQ1), user perceptions of output quality (RQ2), and qualitative feedback.

\begin{table*}[htbp]
    \centering
    \caption{System log events. Latency measures apply only to LLM query events.}
    \Description{Four-column table listing logging categories, event types, explanations, and parameters captured. Two categories. LLM Query Events: new-generation (prompt submission), edit-generation (modify previous prompt submission), re-generation (regenerate response to prior prompt). Parameters logged for query events: EventTarget, Content, Actual time-to-first-token, Timestamp. Note: latency measures apply only to LLM query events. User Action Events: copy (copy output text via highlight or button), paste (paste text into input field), new-chat (start a new AI conversation thread). Parameters logged for user actions: EventTarget, Content, Timestamp.}

    \label{systemlogevents}
    \begin{tabular}{p{2.5cm} p{2.4cm} p{5.1cm} p{4cm}}
    \hline
    \textbf{Category} & \textbf{Event Type} & \textbf{Explanation} & \textbf{Parameters Logged} \\
    \hline
    \multirow{3}{*}{LLM Query Events} 
    & new-generation & Prompt submission 
    & \multirow{3}{=}{\parbox{4cm}{EventTarget, Content, Actual time-to-first-token, Timestamp}} \\
    & edit-generation & Modify previous prompt submission & \\
    & re-generation & Regenerate response to prior prompt & \\
    \hline
    \multirow{3}{*}{User Action Events} 
    & copy & Copy output text (highlight or button) 
    & \multirow{3}{=}{\parbox{4.5cm}{EventTarget, Content, Timestamp}} \\
    & paste & Paste text into input field & \\
    & new-chat & Start a new AI conversation thread & \\
    \hline
    \end{tabular}
    \end{table*}

\subsubsection{Interaction Log Measures}
\label{Measures-InteractionLog}

    Derived from system logs generated by the experimental platform (Section~\ref{System Design}), interaction measures captured session and task identifiers, event type, and event-specific metadata (e.g., text length, source of action).

    Table~\ref{systemlogevents} summarizes the two classes of events: 
    (1) \emph{LLM query events}, which triggered model output and were subject to the latency manipulation, and 
    (2) \emph{user action events}, which reflected participants’ handling of outputs (e.g., copying, pasting) but were not themselves latency-affected. 
    
    \paragraph{Data Preprocessing.}  
    Raw logs were preprocessed before analysis:
    \begin{itemize}[nosep]
       \item \textbf{Deduplication:} Events occurring within a one-second window were flagged as potential duplicates (e.g., accidental double-clicks). In such cases, the first instance was retained; subsequent near-duplicates were removed unless manual inspection indicated they reflected intentional actions.

        \item \textbf{Latency fidelity:} Although fixed latencies (2\,s, 9\,s, 20\,s) were imposed, runtime fluctuations (e.g., network variability) were unavoidable. To preserve treatment fidelity, we computed each participant’s median TTFT across the three tasks and excluded cases where the median exceeded 120\% of the assigned latency. Using a percentage-based cutoff, rather than an absolute time threshold, ensured that participants’ typical waiting experience reflected the intended condition, while the median provided a robust estimate less distorted by outliers. This approach aligns with Weber’s law \cite{haigh_role_2021}, which suggests that time perception scales proportionally rather than absolutely.

    \end{itemize}

    \paragraph{Aggregation.}  
Event counts were aggregated per participant across the three tasks they completed, then analyzed by Latency-Level $\times$ Task-Type. We did not analyze tasks individually because our focus was at the Task-Type level (Creation vs.\ Advice) rather than on unique task variation. This choice is also consistent with our survey design, where post-task ratings were collected once after all three tasks, providing an aggregate judgment of the experience.

\subsubsection{Perceived Response Quality}

    As part of the post-experiment survey, participants rated the AI assistant’s responses on seven items (7-point Likert scale; 1 = Strongly Disagree, 7 = Strongly Agree):  
    
    \begin{itemize}[leftmargin=*]
        \item \textbf{Clarity:} “The AI assistant’s responses were clear and easy to understand.”  
        \item \textbf{Relevance:} “The AI assistant’s responses addressed the task without including irrelevant information.”  
        \item \textbf{Understanding:} “The AI assistant’s responses demonstrated a correct grasp of the task’s meaning and context.”  
        \item \textbf{Thoughtfulness:} “The AI assistant’s responses were thoughtful and took the nuances of the task into account.”  
        \item \textbf{Usefulness:} “The AI assistant’s responses were useful for completing the tasks.”  
        \item \textbf{Trustworthiness:} “I trust the information provided by the AI assistant.”  
        \item \textbf{Expectation Gap:} “The overall quality of the AI assistant’s responses met my expectations.”  
    \end{itemize}
    
    These constructs were selected to capture multiple dimensions of perceived quality, adapted from prior frameworks for evaluating LLM outputs \cite{tam_framework_2024,lee2023evaluating, dillion_ai_2025}. Participants were instructed to base their judgments on their \emph{overall} experience across tasks.

\subsubsection{Cognitive Workload (NASA-TLX)}

Participants also reported their subjective workload via the NASA Task Load Index ~\cite{hart_development_1988} (6 dimensions; 0–100 slider scale): 

\begin{itemize}[leftmargin=*]
    \item \textbf{Mental Demand:} “How mentally demanding were the tasks?”  
    \item \textbf{Physical Demand:} “How physically demanding were the tasks?”  
    \item \textbf{Temporal Demand:} “How hurried or rushed was the pace of the tasks?”  
    \item \textbf{Performance:} “How successful were you in accomplishing what you were asked to do?”  
    \item \textbf{Effort:} “How hard did you have to work to accomplish your level of performance?”  
    \item \textbf{Frustration:} “How insecure, discouraged, irritated, stressed, and annoyed were you?”  
\end{itemize}

\subsubsection{Latency Perception and Qualitative Feedback}

To assess participants’ awareness of delay, the post-experiment survey included the question: \textit{“Did you notice any delay in the AI assistant’s responses while completing the tasks?”} (Yes/No).  

Participants who responded “Yes” ($n=140$) were then displayed an open-ended question: \textit{“In what ways, if any, did the response delays influence how you interacted with the AI Assistant, and how you judged the quality of its responses (e.g., clarity, usefulness, thoughtfulness, trustworthiness), including whether they matched what you expected?”} These reflections were analyzed thematically.

\section{Results} 
\label{Results}

We analyzed the quantitative outcome variables using aligned rank transform (ART) ANOVA \cite{wobbrock_aligned_2011}, given violations of normality confirmed by Shapiro-Wilk tests. Only Latency Perception (categorical) was analyzed using chi-squared test of independence. For significant main effects or interactions, post hoc tests with pairwise comparisons using ART contrasts and Tukey’s HSD adjustment were conducted \cite{lenth_emmeans_2025}.

         \begin{figure}[htbp]
            \centering
            \includegraphics[width=0.46\textwidth]{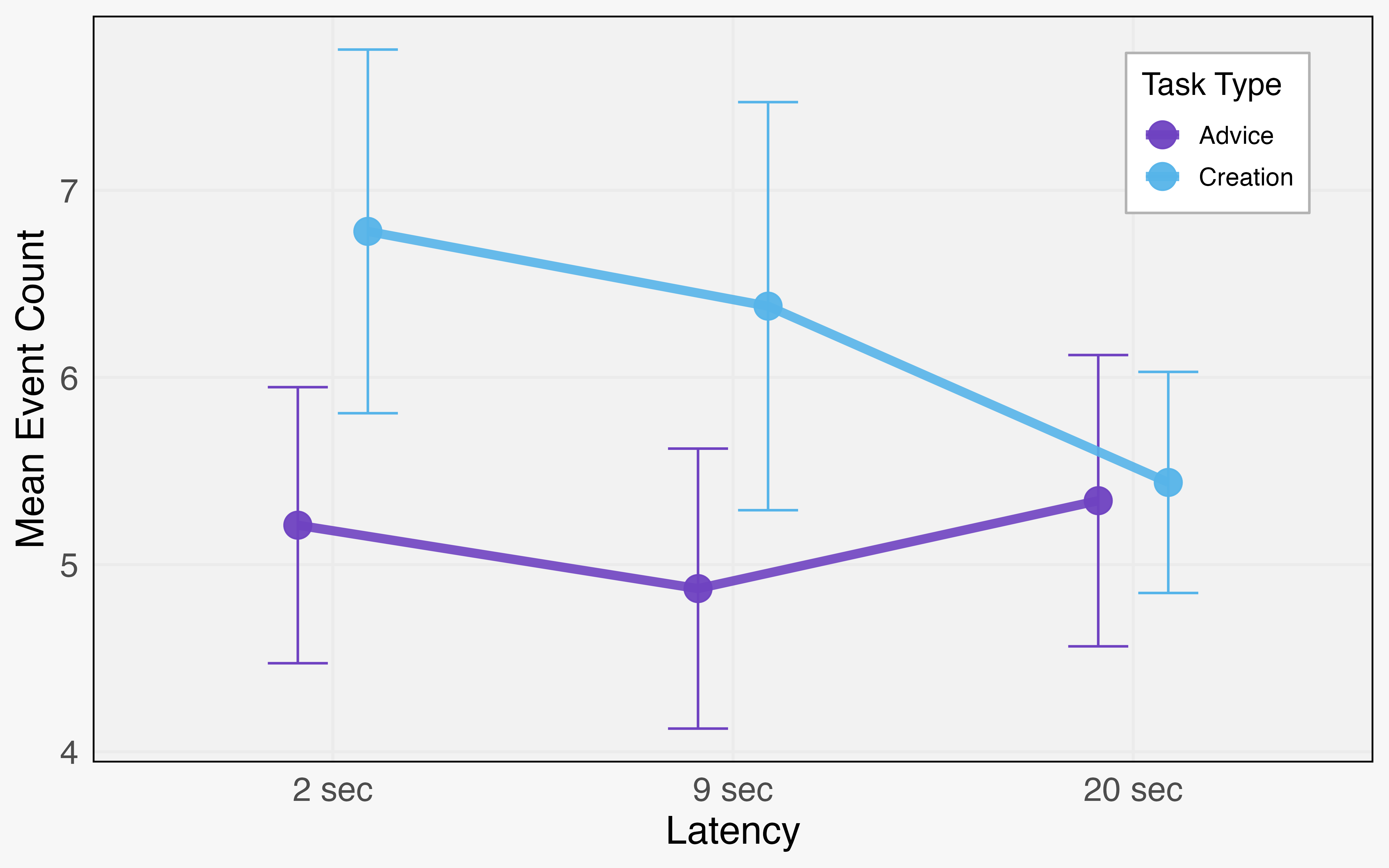}
            \caption{Mean event count per participant for prompt submissions across latency and task type. Participants who completed \emph{Creation} tasks submitted significantly more LLM prompts than those who completed \emph{Advice} tasks. Neither latency level nor the interaction effect was significant, indicating that prompt generation was driven primarily by task type. Error bars represent 95\% confidence intervals.}
            \label{fig:plot_new-prompt}
            
            \Description{Line chart with 95\% confidence intervals showing mean prompt submissions per participant by latency (2, 9, and 20 seconds) and task type. Creation is consistently higher than Advice at each latency. For Creation, means decline with longer latency (about 6.7 at 2 seconds, 6.3 at 9 seconds, 5.4 at 20 seconds). For Advice, means are lower and roughly flat with a slight dip at 9 seconds (about 5.2 at 2 seconds, 4.9 at 9 seconds, 5.3 at 20 seconds). The gap between task types narrows at 20 seconds but remains. Axes: x = Latency (seconds); y = Mean Event Count.}
          \end{figure}

\begin{figure*}[htbp]
      \centering
      \includegraphics[width=0.8\textwidth, trim=0 3 0 10, clip]{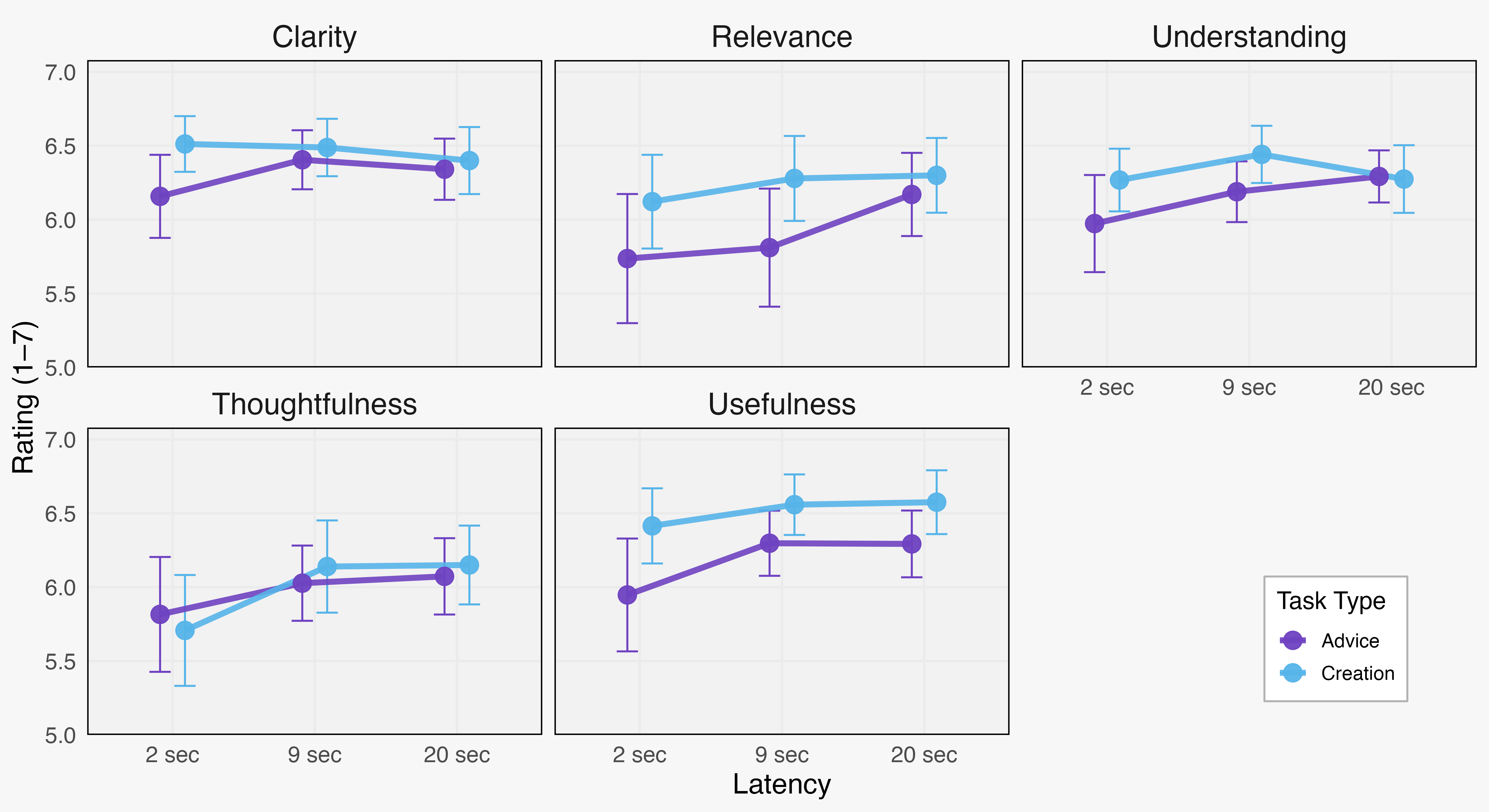}
      \caption{Mean ratings (±95\% CI) by latency and task type. Ratings are displayed on a truncated scale (5–7) for visual clarity; all measures were collected on a 1–7 Likert scale. Creation responses were higher than Advice on Clarity (**), Relevance (*), Understanding (**), and Usefulness (***). Thoughtfulness increased with latency (2\,s < 9\,s ** ; 2\,s < 20\,s *); Usefulness was greater at 9\,s than 2\,s (*). Asterisks denote levels of statistical significance (* $p<.05$, ** $p<.01$, *** $p<.001$).}
      \label{fig:perception_boxplots2}
      
      \Description{Five small-multiple line charts with 95\% confidence intervals show mean ratings (1–7 scale, displayed 5–7) by latency (2, 9, and 20 seconds) and task type (Advice in purple, Creation in blue). Across panels (Clarity, Relevance, Understanding, Thoughtfulness, and Usefulness), Creation ratings are generally higher than Advice at each latency. Thoughtfulness ratings rise across latency levels, and Usefulness is highest at the 9-second condition. Axes show latency on the x-axis and mean rating on the y-axis.}
\end{figure*}

\vspace{-1mm}

\subsection{System Log}
\label{Results-SystemLog}

        \paragraph{\textbf{Model Query Events.}}
        There was a significant main effect of Task-Type on the number of New-Prompt generation events, $F(1,236) = 8.57, p = .004, \eta^2_p = .035$. Participants in the Advice condition ($M = 5.14, SD = 2.33, 95\%$ CI [4.72, 5.57]) submitted fewer new prompts  compared to those in the Creation condition ($M = 6.20, SD = 2.94, 95\%$ CI [5.68, 6.72]). The pairwise comparison was statistically significant, $t(236) = -2.93, p = .004$, Cohen’s $d = -0.40$. Neither Latency-Level nor the interaction effect was significant, suggesting that prompting behavior was driven more by differences in Task-Type than by Latency-Level. 
        
        Editing and regenerating prompts was very rare. Only 13 out of 240 participants (5.4\%) made any edits, while 16 out of 240 participants (6.7\%) regenerated an LLM response. No statistically significant effect was found.

        \paragraph{\textbf{User Action Events.}}
        Copy actions (via text highlight or the copy button) were frequent, with 90.1\% of participants using them at least once. Paste actions were less common (45.8\%), suggesting that while the majority of participants copied model outputs, fewer pasted text outputs directly into the task response box. New-chat resets were rarer, initiated by 10.8\% of participants. For all event types, no significant main or interaction effects were detected ($p> .05$).

    Overall, interaction patterns around LLM prompting were shaped more by Task-Type than by Latency-Level; Creation tasks elicited more New-Prompt submissions. While copying was the most frequent user action, the use of copy, paste, and new-chat features was not affected by condition.

\subsection{Perceived Quality of LLM Responses}
\label{Results-PerceivedQuality}
        
    \paragraph{\textbf{Clarity:}} A significant main effect of Task-Type was found, \\$F(1, 234) = 7.68, p = .006, \eta^2_p = .033$. LLM responses in the Creation condition ($M = 6.47, SD = .64, 95\%$ CI [6.35, 6.58]) were rated clearer than those in the Advice condition ($M = 6.30, SD = .71, 95\%$ CI [6.17, 6.43]). No significant effects of Latency-Level or interaction were observed.
        
    \paragraph{\textbf{Relevance:}} There was a significant main effect of Task-Type, $F(1, 234) = 4.14, p = .043, \eta^2_p = .018$. Participants rated LLM responses in the Creation condition as more relevant ($M = 6.23, SD = .91, 95\%$ CI [6.07, 6.40]) than those in the Advice condition ($M = 5.91, SD = 1.15, 95\%$ CI [5.70, 6.13]). No significant effects of Latency-Level or interaction were observed.
        
    \paragraph{\textbf{Understanding:}} A significant main effect of Task-Type was found, $F(1, 234) = 7.12, p = .008, \eta^2_p = .031$. Participants rated LLM responses in the Creation condition ($M = 6.33, SD = .67, 95\%$ CI [6.21, 6.45]) as demonstrating stronger task understanding than that of the Advice condition ($M = 6.16, SD = .75, 95\%$ CI [6.02, 6.29]). No significant effects of Latency-Level or interaction were observed.
        
    \paragraph{\textbf{Thoughtfulness:}} A significant main effect of Latency-Level emerged, $F(2, 234) = 5.12, p = .007, \eta^2_p = .037$. No significant effect of Task-Type or interaction was detected.  
        
    Post-hoc comparisons indicated that LLM responses at 2\,s latency ($M = 5.76, SD = 1.18, 95\%$ CI [5.50, 6.02]) were rated less thoughtful than those at 9\,s ($M = 6.09, SD = 0.90, 95\%$ CI [5.89, 6.29]; $p = .008$) and 20\,s ($M = 6.11, SD = 0.82, 95\%$ CI [5.93, 6.29]; $p = .040$). No other pairwise differences were significant.  
        
    Overall, responses with longer latencies (9–20\,s) were perceived as more thoughtful compared to very short waits (2\,s).
        
    \paragraph{\textbf{Usefulness:}} Significant main effects of both Task-Type and Latency-Level were observed.  
    For Task-Type, $F(1, 234) = 34.9, p < .001, \eta^2_p = .13$, LLM responses for Creation condition were rated as more useful ($M = 6.52, SD = .72, 95\%$ CI [6.39, 6.64]) than those in the Advice condition ($M = 6.18, SD = .88, 95\%$ CI [6.02, 6.34]; $p < .001$).  For Latency-Level, $F(2, 234) = 3.61, p = .028, \eta^2_p = .032$, post hoc comparisons showed that responses at a 9\,s latency ($M = 6.44, SD = .67, 95\%$ CI [6.29, 6.59]) were rated more useful than those at a 2\,s latency ($M = 6.19, SD = 1.01, 95\%$ CI [5.96, 6.42]; $p = .021$). No other pairwise differences were significant.  
        
    Overall, responses were perceived as more useful in Creation tasks, with usefulness peaking at moderate (9\,s) latency.
        
    \paragraph{\textbf{Trustworthiness:}} No significant main or interaction effects were observed ($p > .05$). Mean ratings across all conditions were moderately high ($M = 5.80$–$6.19$ on a 7-point scale). This suggests participants expressed moderate trust in the LLM’s responses, with little variation conditions.
        
    \paragraph{\textbf{Expectation Gap:}} There were no significant main or interaction effects ($p > .50$). Mean ratings across all conditions were high (6.05–6.42 on a 7-point scale), corresponding to ``agree'' to ``strongly agree.'' This indicates that participants consistently felt the AI Assistant’s responses met their expectations, with little variation across conditions.

\vspace{3mm}
Across measures, participants assigned to Creation tasks perceived the LLM’s responses more favorably than those in Advice tasks, rating them significantly higher in clarity, relevance, understanding, and usefulness. In terms of Latency effects, responses were perceived as more thoughtful and useful at moderate or longer latencies (9–20\,s) compared to very short waits (2\,s).

\subsection{NASA-TLX (Workload)}
\label{Results-NASA-TLX}

    ANOVAs revealed no significant main or interaction effects of Task-Type or Latency-Level on any NASA-TLX dimension (all $p > .05$). Descriptively, participants reported:
    
   \textbf{Mental Demand:} Moderate levels ($M = 52.4$-$56.2$), indicating the tasks required a fair amount of concentration.  

    \textbf{Physical Demand:} Low levels ($M = 20.7$-$26.1$), consistent with the nature of the task.  
    
    \textbf{Temporal Demand:} Moderate levels ($M=25.9$-$34.9$), suggesting participants did not feel too much time pressure.
    
    \textbf{Performance:} High ratings of success ($M = 84.2$-$89.2$), indicating participants generally felt they performed well.  
    
    \textbf{Effort:} Moderate effort ($M = 63.9$-$67.0$), suggesting participants had to work somewhat hard but not excessively so.  
    
    \textbf{Frustration:} Low frustration ($M = 10.3$-$16.9$), suggesting participants rarely felt so.

\subsection{LLM Latency Perception}
\label{Results-LatencyPerception}

        \begin{figure}[htbp]
          \centering
          \includegraphics[width=0.45\textwidth]{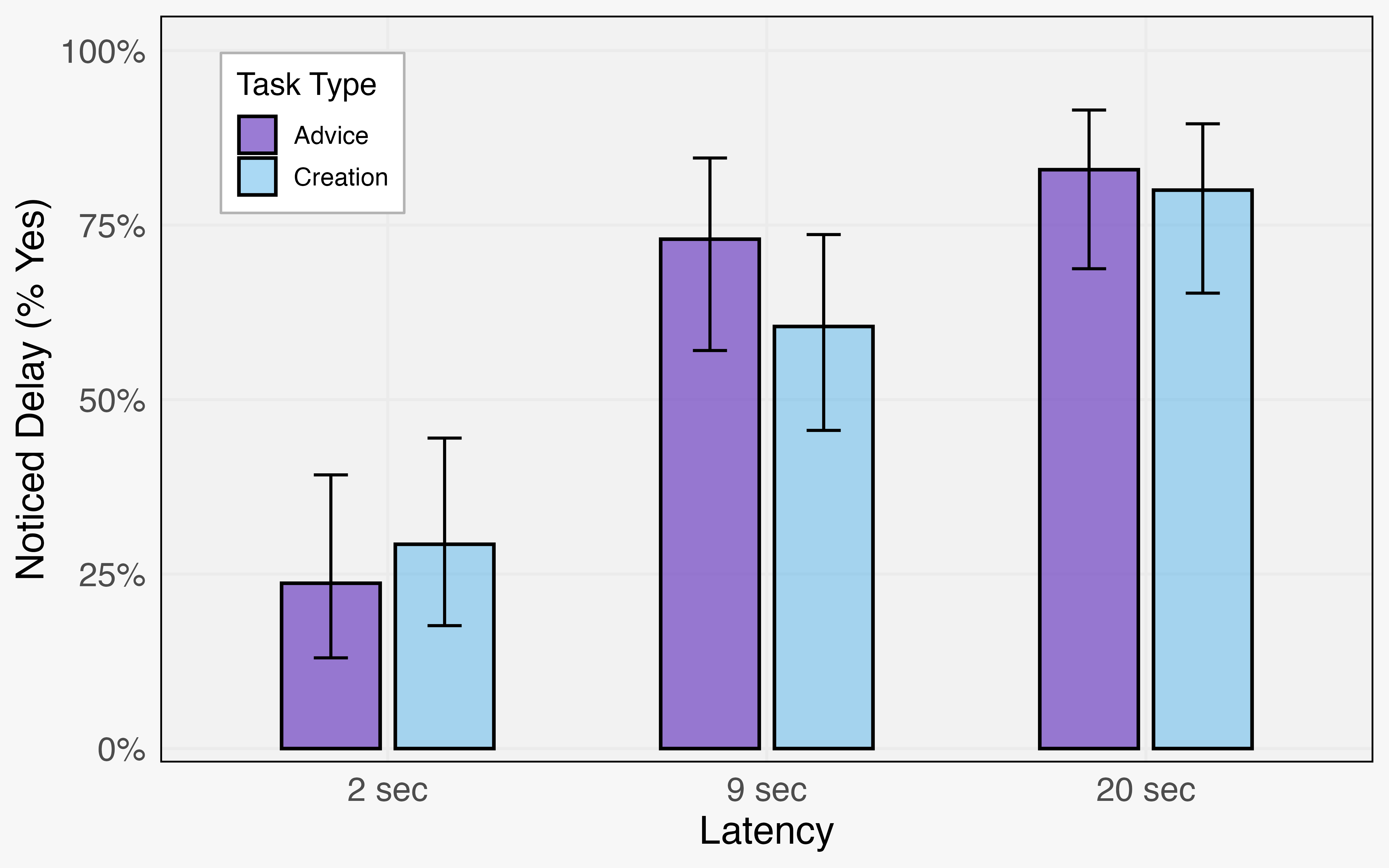}
          \caption{Percentage of participants who reported noticing the LLM’s response delay (“Yes”) by latency level and task type. Bars show proportions with Wilson 95\% CIs.}
          \label{fig:latencyrecall}
          
          \Description{Grouped bar chart with Wilson 95\% confidence intervals showing the percentage of participants who noticed the LLM’s delay by latency (2, 9, 20 seconds) and task type (Advice purple, Creation blue). Detection rises sharply with latency: about 25–30\% at 2 seconds, around 60–75\% at 9 seconds, and roughly 80–85\% at 20 seconds. Advice is slightly higher than Creation at 9 and 20 seconds, while Creation is slightly higher at 2 seconds.}

        \end{figure}

    A chi-square test of independence revealed a strong association between condition and latency perception, $\chi^2(5, N = 240) = 54.29, p < .001$, Cramer’s $V = .48$, indicating a large effect. In the Creation task, 29.3\% noticed latency at 2\,s, rising to 60.5\% at 9\,s, and 80.0\% at 20\,s. A similar trend was observed in the Advice task, with 23.7\% noticing latency at 2\,s, 73.0\% at 9\,s, and 82.9\% at 20\,s. This pattern suggests that participants were more likely to perceive latency at longer wait times, regardless of task type.

\subsection{Qualitative Response}
\label{Results-Qualitative}

    Only participants who reported noticing a delay (in the Latency Perception question) provided open-ended reflections ($n=140$). We thematically analyzed open-ended responses to characterize how participants interpreted the delay. The first author created an initial codebook based on a pilot set. Two non-author coders independently coded the full dataset; any discrepancies were resolved by discussion. The inter-rater reliability was $\kappa=.46$--$.80$, all $p<.001$.

    Definitions and illustrative quotes are provided in Appendix~\ref{appendix-Qualitative themes} to clarify and ground each theme. Below, we report theme counts to convey relative salience rather than to test between-group differences.

    \paragraph{\textbf{Theme 1: No/Minimal Impact.}}  
    This was the most frequent theme overall, coded for 63 participants ($\approx$45\%). Responses indicated that while delays were noticed, they were often regarded as inconsequential to how participants interacted with or evaluated the AI. Within-condition proportions suggest this framing was especially common in the shortest waits (44-58\% across 2\,s conditions, compared to $\approx$37–47\% across the 9\,s and 20\,s conditions). The pattern was similar across task types.

    \paragraph{\textbf{Theme 2: Inferred AI Deliberation.}}  
    43 participants ($\approx$31\%) interpreted the latency as evidence of the AI “thinking” or “processing,” sometimes raising expectations for more thoughtful or higher-quality outputs. This interpretation was more typical of longer pauses, suggesting they were more readily framed as constructive deliberation. Prevalence was similar across task types. 
    
    \paragraph{\textbf{Theme 3: Quality-Contingent Value.}}  
    14\% of the participants emphasized that waiting was acceptable if the LLM ultimately produced clear, detailed, or accurate responses. The perceived worth of waiting was thus conditional on output quality. A notable dip occurred in the 20\,s Advice condition ($\approx$6\%), suggesting that very long waits during Advice tasks were less perceived as worthwhile.

    \paragraph{\textbf{Theme 4: Self-Regulatory Use of Wait Time.}}  
    Only 5 participants ($\approx$4\%), all in the 9\,s latency conditions, described using delays productively (e.g., planning the next query or re-reading instructions).

    \paragraph{\textbf{Theme 5: Negative Affective Cost.}}  
    Only 7\% of the participants described delays as frustrating, anxiety-inducing, or disruptive. Mentions of this theme rose as latencies extended.
    
    \paragraph{\textbf{Theme 6: System Reliability Concern.}}  
    14\% of all participants attributed delays to potential technical or system issues (e.g., poor connectivity or model inefficiency). It was most prevalent in the longest waits and Advice tasks.
    
    \paragraph{\textbf{Theme 7: Interaction Strategy Adaptation.}}  
    9\% of all participants reported adapting their prompting strategies in response to delays; for example, consolidating more information into single prompts, reducing the number of queries, or double-checking outputs. Such adaptations emerged primarily at the higher latencies, indicating that longer delays encouraged more strategic adjustments to interaction style. 

\vspace{4mm}

Taken together, these themes reveal a spectrum of ways participants made sense of latency, from dismissing it as inconsequential to their task workflow, to reframing it as evidence of deliberation or a chance for productive use. We use these mechanisms to ground the design discussion and implications that follow.

  \section{Discussion}
\label{Discussion}

This study investigated how response latency and task type shape users’ interaction with LLMs. The following sections interpret the findings, connecting them to the broader HCI literature and extracting implications for design, ethics, and future research.

\subsection{Interaction Behavior: Robustness to Latency (RQ1)}

    \paragraph{Consistent Behaviors Across Latency Durations.}  
    
    Regardless of whether onset delays were 2\,s, 9\,s, or 20\,s, participants submitted new prompts, copied outputs, and pasted or refreshed the chat at similar rates. Editing and regenerating were rare overall ($\approx$5–7\% of participants engaged in such actions), suggesting a strong preference for issuing new prompts over modifying or re-running prior ones. The minimal revisiting of previous responses aligns with evidence that modifying earlier AI outputs within chat-based interfaces imposes substantial interaction costs: users must scroll through transcripts, locate specific elements in prior responses, and manually reference them, making such revision comparatively effortful \cite{gibbons_accordion_2023}. Beyond interactional frictions, motivational factors may also contribute. XAI research shows that users choose to invest effort in additional interaction only when they expect it to advance their goals \cite{liao_questioning_2020}. Notably, this pattern held even at the longest delays (20\,s). Participants did not prefer more conservative strategies to avoid the “cost” of waiting. Instead, they consistently favored new prompts.

    In response to the question, \textit{“Did you notice any delay in the AI assistant’s response while completing the tasks?”}, recognition rose significantly with latency level, from $\approx$26.6\% at 2\,s to 66.3\% at 9\,s and 81.5\% at 20\,s (averaged across task types), confirming the effectiveness of the experimental manipulation. Several interpretations may account for why some participants, particularly in longer latency conditions, did not report noticing the wait.
    Users may normalize pauses as conversational turn-taking \cite{sacks_simplest_1974, hwang_when_2019}, interpret the pause as the AI “thinking” \cite{kuang_enhancing_2024} rather than as a delay, pay more conscious attention to the content of the output than to its timing, or inaccurately recall the delay \cite{ghafurian_countdown_2020}. These tendencies align with broader patterns in how users make sense of waiting, which we elaborate in the next section.

     \paragraph{Latency as Normalized Waiting.}  
    One explanatory lens for this robustness in interaction behavior is the conversational structure of LLM interaction. Conversation analyses show that pauses are a routine and interpretable feature of turn-taking~\cite{sacks_simplest_1974}. In LLM-based knowledge work, pauses are normalized as part of a turn-based exchange with a “black box” partner, and Kuang et al.~\cite{kuang_enhancing_2024} highlight how timing in conversational AI shapes user trust and engagement.

     Furthermore, qualitative feedback in our study indicates that the majority of participants either downplayed delays as having \emph{No or Minimal Impact} ($\approx$45\%) or reinterpreted them as signs of \emph{AI Deliberation} ($\approx$31\%). As P23 explained, \emph{“I just assumed since the AI was providing quite a bit of information, that it was taking a little time to generate”} (20\,s, Advice). This pattern suggests that participants not only normalized waiting but in some cases anthropomorphized the system, attributing delays to the AI “thinking” and even showing a degree of empathy toward its effort \cite{araujo_living_2018}.
     
     At the same time, a smaller subset of participants reported adapting their interaction strategies to avoid repeated exchanges. P1 shared: “I included as much information I could with the prompt, knowing that if I did not I would have to follow up with another prompt, which would make the task even longer” (20\,s, Creation), and P20 stated, “I tried to minimize the number of prompts per task because I got annoyed at the response delay” (20\,s, Advice). Of the 12 participants who described changing their behavior in this way, 7 were in the 20\,s condition. While not conclusive, this pattern may reflect a satisficing tendency \cite{dang_choice_2023} at higher latencies, i.e. when an answer was usable, participants often reduced further prompting rather than continuing to iterate. Evidence of other costs, such as frustration or concerns about reliability, also became more common at the longest 20\,s delays. While these experiences were not widespread enough to alter overall interaction patterns, they highlight boundary conditions under which latency may begin to undermine engagement.

\subsection{Prompting Frequency: Shaped by Task Type (RQ1)}

\paragraph{Cognitive Demands of Creation vs.\ Advice Tasks.}

A clear effect of \emph{task type} emerged in prompting behavior. Participants attempting Creation tasks submitted significantly more new prompts ($M=6.20$, $SD=2.94$) than those in Advice tasks ($M=5.14$, $SD=2.33$). Other logged user actions (e.g., copy–paste) showed no systematic differences (Section~\ref{Results-SystemLog}), underscoring that the main behavioral divergence lay in prompt submissions.

This pattern aligns with the divergent–convergent split in knowledge work. \emph{Divergent} tasks, i.e., open-ended exploration, often encourage repeated prompting, as users probe the model to expand or refine ideas. For example, a task like slogan generation (in the Creation condition) could plausibly invite participants to request multiple variations until one “felt right.” By contrast, evaluative Advice tasks may conclude once a satisfactory answer is obtained. Put differently, Creation tasks encourage prompting because each turn can produce new ideas, whereas in Advice tasks the marginal value of further queries diminishes after sufficiency is reached. Prior work similarly shows that generative AI sustains divergent cycles of idea generation~\cite{brynjolfsson_generative_2025,lee_impact_2025}, whereas convergent use curtails further interaction once sufficiency is reached~\cite{liu_how_2024,brachman_current_2025,liao_questioning_2020}.

\paragraph{Design and engineering implications.}  
The significant influence of task-type on prompting frequency may carry practical consequences. For developers of LLM-based applications, Creation tasks (which elicited more prompts) are likely to entail greater token consumption, longer interaction chains, and increased context-retention overhead, thereby increasing computational costs and infrastructure demands. By contrast, Advice tasks may generate shorter sessions with fewer turns, but appear more sensitive to the impact of individual delays. These distinctions offer a starting point for reasoning about how task orientation structures interaction patterns and how such differences may be accounted for in system design, optimization, and resource planning.

\subsection{User Perceptions: Latency’s Influence on Quality and Experience (RQ2)}

   Perceptions of AI output quality are critical because they shape trust, reliance, and willingness to integrate results into knowledge work~\cite{lee_while_2025,liao_questioning_2020}. Prior research shows that subjective impressions are as crucial as objective accuracy in determining adoption and use~\cite{davenport_thinking_2005,brynjolfsson_generative_2025}. In our study, all LLM responses were generated by the same underlying model, yet variations in latency and task type significantly shaped how users judged their quality.
    
\paragraph{\textbf{Perceived Thoughtfulness.}}
Responses delivered after very short delays (2\,s) were judged as less thoughtful than those following moderate or longer waits (9–20\,s). This pattern aligns with the \emph{AI Deliberation} theme, in which slower responses were interpreted as evidence of system deliberation. Such interpretations are consistent with the effort heuristic, whereby people infer quality from perceived effort~\cite{kruger_effort_2004}, and, in this context, are accompanied by anthropomorphic attributions of deliberation to the system. Prior work similarly shows that slower algorithmic advice can be judged as more accurate or trustworthy because delays are read as signs of deliberation~\cite{park_slow_2019}. Overall, this suggests that latency can bias evaluations of depth independently of content.

\paragraph{\textbf{Perceived Usefulness.}}
Ratings were highest at moderate latencies ($\approx$9\,s) and lower at both shorter (2\,s) and longer (20\,s) waits, and were significantly higher for Creation tasks than for Advice tasks. Unlike thoughtfulness, which is closely associated with users’ attributions of cognitive effort, usefulness reflects the pragmatic value of the output for accomplishing the task, grounding evaluation in instrumental utility rather than process attributions~\cite{tam_framework_2024}. Consistent with this view, prior work on chatbot guidance shows that the same content can be judged as useful or disruptive depending on when it is delivered~\cite{yeh_how_2022}, underscoring the temporal contingency of perceived utility. For design, this suggests a trade-off: moderate pauses may enhance perceived usefulness, whereas longer delays risk inflating expectations, particularly in Advice contexts where additional “thinking time” does not necessarily yield better outputs~\cite{sui_stop_2025}. Designers must therefore weigh the potential benefits of calibrated pauses against the risk of overstating the practical value of the system’s outputs.

    \paragraph{\textbf{Trust and Acceptance vs.\ ``Too Slow''.}}  
    Trust ratings remained high across all latency conditions, even at 20\,s delays. Expectation-gap scores similarly indicated that participants, regardless of task type and latency level, felt the AI assistant’s responses met their expectations. Qualitative accounts reinforce this pattern: many participants expressed a \emph{quality-contingent value} perspective, treating delays as acceptable, or even “worth it” (P129, P135), when the output was clear, detailed, or accurate.  For instance, P104 remarked \emph{``I did not mind the response delays due to the quality of the responses''} (9\,s, Creation). In these cases, participants granted the slower AI the benefit of the doubt so long as the response itself was satisfactory.
\vspace{2mm}

    However, qualitative feedback also revealed limits to this tolerance. In the longest-delay condition (20\,s), especially for Advice tasks, some participants began to lose patience or question the system's reliability: ``The responses were fine but the response time made me question if my internet cut off'' (P100, 20\,s, Advice). Perceived usefulness dipped slightly in these cases, and explicit mentions of waiting being ``worth it'' declined. This suggests a threshold effect: modest delays can enhance perceived value, but very long waits risk backfiring, reframing latency as inefficiency rather than deliberation.

    \paragraph{\textbf{Calibrating the “Sweet Spot” of Latency.}}  
    Across measures, the findings suggest a temporal zone of benefit. Very fast responses (2\,s) were often discounted as superficial, while moderate latencies (9\,s) produced the most favorable evaluations, particularly for usefulness. Longer waits (20 s) did not enhance thoughtfulness and, for some participants, tipped into frustration or reliability concerns. In other words, modest pauses could enhance perceptions of quality, but excessive waiting risked backfiring, reframing latency as inefficiency rather than deliberation. This aligns with the principle of \emph{positive friction}~\cite{chen_exploring_2024,inan_better_2025}, which holds that small, well-calibrated slowdowns can foster reflection, whereas excessive friction undermines engagement.

\vspace{2mm}
    Crucially, these perceptual shifts were not mirrored in workload outcomes. No NASA-TLX dimension (e.g., frustration, effort, performance) showed significant differences across latency levels or task type. This dissociation suggests that latency shaped how users \emph{interpreted} the LLM responses, not how much cognitive burden they experienced. Some participants even qualitatively described using pauses productively (e.g., reviewing instructions, planning the next query, or briefly multitasking), folding latency into the rhythm of their workflow. From a distributed cognition perspective~\cite{Hollan2000DistributedC}, such appropriation illustrates how temporality is not just a single system's parameter but a cognitive resource that users coordinate across tasks, artifacts, and activities.

\vspace{2mm}

As LLMs deliver increasingly fast responses, users’ interpretations of latency may recalibrate with repeated exposure and more accurate mental models of system capability. Such recalibration may weaken attributions of AI “deliberation,” reducing the tendency to view fast responses as shallow or mismatched to task demands. At the same time, users may remain sensitive to delays that violate learned expectations. Understanding how temporal expectations evolve under sustained use is therefore an important direction for longitudinal research.

\subsection{Design Implications} 

While prior HCI research has treated latency largely as a model performance constraint, our results show that it also shapes meaning-making in interaction. This reframing expands the design space: rather than optimizing solely for speed, designers can intentionally use response timing to influence perceived quality, trust, and engagement. Below, we outline four design implications.

\begin{itemize}
    \item \textbf{Instantaneous output need not be the priority.}  
    In our study, the shortest onset delay (2\,s) was consistently rated less thoughtful and less useful than slower responses. This suggests that faster replies can undercut perceptions of depth. Developers of productivity tools (e.g., writing assistants or research copilots) might deliberately pace responses, even if the model could respond faster.
    
    \item \textbf{Mid-length pauses can be leveraged strategically.}  
    Moderate waits ($\approx$9\,s) were occasionally reframed by participants as useful, with some reporting they used the time to re-read instructions or plan next steps. This suggests that latency can be repurposed as a resource. In learning technologies, such pauses might be paired with micro-prompts (e.g., ``What would you expect the answer to include?'') to engage users in anticipatory thinking. Lightweight system cues (``analyzing your request…'') can also normalize waiting and preserve trust. Section \ref{Limitations and Future Work} suggests testing whether scaffolds during waits amplify their reflective value.
    
    \item \textbf{Mitigate the costs of long delays.}  
    Even at 20\,s, latency did not substantially alter prompting behavior, though a subset of participants reported concerns about system reliability. These reactions suggest a potential boundary condition: very long delays risk being interpreted as inefficiency rather than deliberation. These risks may be especially salient in contexts where timely feedback matters.

    \item \textbf{Calibrate latency to task orientation.}  
    Creation tasks elicited more prompting than Advice tasks, and their outputs were rated more favorably across several quality dimensions. This suggests that latency design should be task-aware rather than uniform. A practical implication is that systems could adapt response pacing based on inferred task intent, treating latency as a tunable interaction parameter rather than a fixed system property.
\end{itemize}

\subsection{Ethical Considerations}
    Our findings highlight that latency is not only a technical parameter but also an interpretive cue with ethical implications. Users in our study often read longer delays as signs of deliberation, even though the underlying content was generated by the same AI model. Such interpretations risk uneven effects across user groups - novices or those with less domain knowledge may be more likely to over-trust “slow” outputs, while experts remain more skeptical ~\cite{gnewuch_chatbot_2018,liao_designing_2022}. This raises concerns about fairness and equity, as latency could inadvertently reinforce disparities in how different populations calibrate trust in AI systems. 
    
    The ethics of latency are also highly contextual. In reflective or creative work, modest delays may provide valuable pauses for sensemaking, while in time-critical settings such as healthcare, finance, or safety, even small artificial waits could introduce risk or erode trust. Designers should therefore avoid one-size-fits-all approaches and consider whether adaptive or user-controlled pacing better supports diverse situational needs.

   At the same time, deliberately manipulating response speed sits uneasily between positive friction and potential deception. While prior work shows that modest delays can reduce automation bias and promote reflection~\cite{bucinca_trust_2021,park_slow_2019}, the absence of clear explanations for timing can lower perceived transparency and trust~\cite{zhang_explaining_2024}. Recent work on deceptive design patterns in LLM-based interfaces sharpens this concern. Shi’s taxonomy~\cite{shi_siren_2025} identifies how temporal cues (such as hesitation, slowed pacing, or staged disclosure of output) can function as \emph{performative deliberation}. This deceptive interaction pattern simulates cognitive effort and leads users to overattribute reasoning or competence to the system.

    This tension is well-captured by discussions of \emph{benevolent deception}~\cite{adar_benevolent_2013}, which note that designers sometimes introduce deceptive cues to bridge gaps between system behavior and user expectations. The Motive–Means–Opportunity model provides an analytic vocabulary for examining when such practices arise. In the case of latency manipulation, the motive may be to support reflection or reduce human error, the means involve shaping observable system behavior, and the opportunity arises because users cannot readily verify true processing time. Crucially, the ethical status of such interventions is context-dependent, shaped by the availability of truthful alternatives and long-term implications for user trust~\cite{adar_benevolent_2013}.

    These considerations underscore the importance of transparency when timing is intentionally shaped. Delays arising from genuine computation may require little justification, but when pacing is designed to influence user interpretation, offering users insight into this behavior, or control over it, helps preserve agency. Without such transparency, users may conflate interactional pacing with signals of epistemic depth, leading to distorted trust calibration.

\section{Limitations and Future Work}
\label{Limitations and Future Work}

\paragraph{\textbf{Latency Manipulation.}} 
Our study manipulated only time-to-first-token (TTFT), the onset of latency, while holding streaming throughput constant and testing three discrete latency levels (2, 9, and 20\,s). TTFT captures an important component of perceived responsiveness, but it represents only one part of the temporal experience. Other factors (such as streaming pace, total response duration, variability across turns, and the relation between latency and output length) likely also shape how latency is perceived. Accordingly, our findings can only support comparative claims across coarse latency levels, but do not establish whether user perceptions follow a continuous curve or hinge on discrete temporal thresholds. Future work should therefore report full timing schematics (assigned and observed TTFT, token counts, total generation time) and test finer-grained manipulations, including relative changes in TTFT (Weber-like proportionality), throughput, and cue-based transparency.

\paragraph{\textbf{Task Context and Generalizability.}}
Our study was conducted in a controlled experimental environment, where participants faced no real consequences for latency and may therefore have been more tolerant of waiting than in professional or time-critical contexts. This tolerance likely reflects the absence of external pressures such as deadlines, accountability to colleagues, or financial stakes. Moreover, the tasks were designed so that all required information was presented within a single interface panel (Figure~\ref{fig:taskinterface}), minimizing the need to retrieve, coordinate, or synthesize information across digital, physical, interpersonal, or cognitive sources. By contrast, real-world knowledge work is typically distributed and interleaved: workers move between files, applications, ideas, and conversations under competing demands. In such settings, latency may be absorbed into opportunities for multitasking or, alternatively, disrupt the coordination required across people, artifacts, and contexts. Future work should investigate latency in ecologically richer settings (e.g., multi-document tasks, collaborative workflows, time-sensitive decisions) to understand how delays shape broader work rhythms.

\paragraph{\textbf{Sociocultural Context.}}  
Participants were recruited from Prolific and limited to U.S.-based adults who reported regular use of AI assistants, implying relatively high digital familiarity and exposure to LLM systems. However, prior work suggests that cultural norms and digital literacy shape how waiting is experienced, influencing patience thresholds, tolerance for uncertainty, and trust in system feedback \cite{maister_psychology_1985}. Expectations of immediacy versus deliberation may therefore differ across cultures and between novice and expert users. Future work should examine response latency in more diverse cultural and demographic contexts, to understand how perceptions of latency, quality, and trust in LLMs vary across groups.

\paragraph{\textbf{Generation Cues and Transparency Interventions.}}
We deliberately adopted a cue-free interface, omitting progress indicators such as ``thinking'' labels during the study's wait period. This design choice allowed us to isolate whether time itself could act as a signal to users, independent of any UI messaging. In practice, however, some commercial LLM platforms surface feedback during waiting, ranging from simple progress indicators to emerging ``reasoning-mode'' displays that reveal intermediate steps of model processing. Such cues may influence not only perceived thoughtfulness and trust but also how users structure their prompting strategies and workflow transitions. Future work can systematically vary cue presence, type, and granularity alongside latency to better understand how temporal signals and transparency jointly shape perceptions of effort and trust.

\paragraph{\textbf{Latency and the Future of Work.}}  
Future work should extend these insights beyond isolated prompt–response exchanges to more complex, agentic scenarios. In emerging forms of delegated or autonomous AI, response times may scale dramatically or take new forms (e.g., background tasks that return results minutes or hours later). In such contexts, the significance of latency may depend less on its duration than on how delays are surfaced, scheduled, or integrated into broader workflows.

\emph{Productive use of pauses.}  
Some participants reported using pauses to review instructions, plan the next query, or briefly multitask. This suggests that waiting may function as a resource for coordination, not merely an interruption. Future work should examine when and how users appropriate pauses across different types of work, and under what conditions delays support reflection versus distraction.

\emph{Synchronous vs.\ asynchronous collaboration.}  
As AI systems evolve from reactive chat tools to semi-autonomous collaborators, the experience of waiting may be reconfigured. In asynchronous scenarios (such as concurrent human work and background AI processing), latency may be absorbed as users shift attention elsewhere and receive results just in time. By contrast, in synchronous collaboration (e.g., real-time decision support), even small delays may remain disruptive if they slow turn-taking or coordination. Exploring how latency design should differ across synchronous and asynchronous settings is an important direction for future research.

\section{Conclusion}

This study examined how LLM response latency and task type shape user interaction. We found that user interaction behaviors remained consistent across latency conditions, but perceptions of output quality varied. Short delays reduced perceived thoughtfulness and usefulness, moderate delays improved evaluations, and long delays sometimes undermined perceptions of system reliability. 

From a design perspective, our findings suggest that response speed should be \emph{calibrated} rather than minimized, since different task contexts demand different balances between immediacy, perceived effort, and reliability. Building on this, a social perspective highlights how users sometimes humanize the AI system itself, interpreting delays as signs of deliberation and even responding with patience or empathy. These social dynamics, in turn, surface ethical concerns: interpretations of delay may calibrate trust in unequal ways across user groups, raising questions about fairness and transparency. Finally, from a future research perspective, these insights point toward the need to examine latency in more complex and distributed work settings, where the rhythms of waiting may align with or disrupt AI-supported knowledge work.

Taken together, this work positions latency as a central dimension of interaction design for AI systems: not simply a barrier to be removed, but a design factor that shapes perceptions, trust, and collaboration.

\begin{acks}
 This work was supported by the National Science Foundation (NSF) under grants 1928614 and 2129076. We are also grateful to our colleagues for their constructive feedback.
\end{acks}

\bibliographystyle{ACM-Reference-Format}
\bibliography{references}

@misc{shi_siren_2025,
	title = {The {Siren} {Song} of {LLMs}: {How} {Users} {Perceive} and {Respond} to {Dark} {Patterns} in {Large} {Language} {Models}},
	shorttitle = {The {Siren} {Song} of {LLMs}},
	url = {http://arxiv.org/abs/2509.10830},
	doi = {10.48550/arXiv.2509.10830},
	
	urldate = {2025-12-04},
	publisher = {arXiv},
	author = {Shi, Yike and Xiao, Qing and Hu, Qing and Shen, Hong and Shen, Hua},
	month = sep,
	year = {2025},
	
	}

@inproceedings{adar_benevolent_2013,
	address = {New York, NY, USA},
	series = {{CHI} '13},
	title = {Benevolent deception in human computer interaction},
	isbn = {978-1-4503-1899-0},
	url = {https://doi.org/10.1145/2470654.2466246},
	doi = {10.1145/2470654.2466246},
	
	urldate = {2025-12-03},
	booktitle = {Proceedings of the {SIGCHI} {Conference} on {Human} {Factors} in {Computing} {Systems}},
	publisher = {Association for Computing Machinery},
	author = {Adar, Eytan and Tan, Desney S. and Teevan, Jaime},
	month = apr,
	year = {2013},
	pages = {1863--1872},
}

@misc{gibbons_accordion_2023,
	title = {Accordion {Editing} and {Apple} {Picking}: {Early} {Generative}-{AI} {User} {Behaviors}},
	shorttitle = {Accordion {Editing} and {Apple} {Picking}},
	url = {https://www.nngroup.com/articles/accordion-editing-apple-picking/},
	
	language = {en},
	urldate = {2025-12-01},
	journal = {Nielsen Norman Group},
	author = {Gibbons, Sarah and Mugunthan, Tarun and Nielsen, Jakob},
	month = sep,
	year = {2023},
}

@article{ghafurian_countdown_2020,
	title = {Countdown timer speed: {A} trade-off between delay duration perception and recall},
	volume = {27},
	issn = {1557-7325},
	shorttitle = {Countdown timer speed},
	doi = {10.1145/3380961},
	
	number = {2},
	journal = {ACM Transactions on Computer-Human Interaction},
	author = {Ghafurian, Moojan and Reitter, David and Ritter, Frank E.},
	year = {2020},
	
	pages = {1--25},
}

@article{stanovich_individual_2000,
	title = {Individual differences in reasoning: implications for the rationality debate?},
	volume = {23},
	issn = {0140-525X},
	shorttitle = {Individual differences in reasoning},
	doi = {10.1017/s0140525x00003435},
	
	language = {eng},
	number = {5},
	journal = {The Behavioral and Brain Sciences},
	author = {Stanovich, K. E. and West, R. F.},
	month = oct,
	year = {2000},
	pmid = {11301544},
	pages = {645--665; discussion 665--726},
}

@book{kahneman_thinking_2011,
	title = {Thinking, {Fast} and {Slow}},
	isbn = {978-1-4299-6935-2},
	publisher = {Farrar, Straus and Giroux},
	author = {Kahneman, D.},
	year = {2011},
	lccn = {2011027143},
}

@article{Hollan2000DistributedC,
  title={Distributed cognition},
  author={James Hollan and Edwin L. Hutchins and David Kirsh},
  journal={ACM Transactions on Computer-Human Interaction (TOCHI)},
  year={2000},
  volume={7},
  pages={174-196},
  doi = {10.1145/353485.353487},
}

@book{nielsen_usability_1994,
	address = {San Francisco, CA, USA},
	title = {Usability {Engineering}},
	isbn = {978-0-08-052029-2},
	
	publisher = {Morgan Kaufmann Publishers Inc.},
	author = {Nielsen, Jakob},
	month = oct,
	year = {1994},
}

@inproceedings{arapakis_impact_2014,
	address = {New York, NY, USA},
	series = {{SIGIR} '14},
	title = {Impact of response latency on user behavior in web search},
	isbn = {978-1-4503-2257-7},
	url = {https://doi.org/10.1145/2600428.2609627},
	doi = {10.1145/2600428.2609627},
	
	urldate = {2025-01-24},
	booktitle = {Proceedings of the 37th international {ACM} {SIGIR} conference on {Research} \& development in information retrieval},
	publisher = {Association for Computing Machinery},
	author = {Arapakis, Ioannis and Bai, Xiao and Cambazoglu, B. Barla},
	month = jul,
	year = {2014},
	pages = {103--112},
}

@inproceedings{lee_impact_2025,
    author = {Lee, Hao-Ping (Hank) and Sarkar, Advait and Tankelevitch, Lev and Drosos, Ian and Rintel, Sean and Banks, Richard and Wilson, Nicholas},
    title = {The Impact of Generative AI on Critical Thinking: Self-Reported Reductions in Cognitive Effort and Confidence Effects From a Survey of Knowledge Workers},
    year = {2025},
    isbn = {9798400713941},
    publisher = {Association for Computing Machinery},
    address = {New York, NY, USA},
    url = {https://doi.org/10.1145/3706598.3713778},
    doi = {10.1145/3706598.3713778},
    booktitle = {Proceedings of the 2025 CHI Conference on Human Factors in Computing Systems},
    articleno = {1121},
    numpages = {22},
    series = {CHI '25}
}

@inproceedings{brachman_how_2024,
	address = {New York, NY, USA},
	series = {{CHI} {EA} '24},
	title = {How {Knowledge} {Workers} {Use} and {Want} to {Use} {LLMs} in an {Enterprise} {Context}},
	isbn = {979-8-4007-0331-7},
	url = {https://dl.acm.org/doi/10.1145/3613905.3650841},
	doi = {10.1145/3613905.3650841},
	
	urldate = {2025-04-06},
	booktitle = {Extended {Abstracts} of the {CHI} {Conference} on {Human} {Factors} in {Computing} {Systems}},
	publisher = {Association for Computing Machinery},
	author = {Brachman, Michelle and El-Ashry, Amina and Dugan, Casey and Geyer, Werner},
	month = may,
	year = {2024},
	pages = {1--8},
	 }

@inproceedings{miller_response_1968,
	address = {San Francisco, California},
	title = {Response time in man-computer conversational transactions},
	url = {http://portal.acm.org/citation.cfm?doid=1476589.1476628},
	doi = {10.1145/1476589.1476628},
	language = {en},
	urldate = {2025-04-09},
	booktitle = {Proceedings of the {December} 9-11, 1968, {AFIPS} '68 ({Fall}, part {I})},
	publisher = {ACM Press},
	author = {Miller, Robert B.},
	year = {1968},
	pages = {267},
	 }

@article{brynjolfsson_generative_2025,
	title = {Generative {AI} at {Work}},
	volume = {140},
	issn = {0033-5533},
	url = {https://doi.org/10.1093/qje/qjae044},
	doi = {10.1093/qje/qjae044},
	
	number = {2},
	urldate = {2025-04-18},
	journal = {The Quarterly Journal of Economics},
	author = {Brynjolfsson, Erik and Li, Danielle and Raymond, Lindsey},
	month = may,
	year = {2025},
	pages = {889--942},
	 }

@inproceedings{liu_how_2024,
	address = {New York, NY, USA},
	series = {{CHI} '24},
	title = {How {AI} {Processing} {Delays} {Foster} {Creativity}: {Exploring} {Research} {Question} {Co}-{Creation} with an {LLM}-based {Agent}},
	isbn = {979-8-4007-0330-0},
	shorttitle = {How {AI} {Processing} {Delays} {Foster} {Creativity}},
	url = {https://dl.acm.org/doi/10.1145/3613904.3642698},
	doi = {10.1145/3613904.3642698},
	
	urldate = {2025-04-27},
	booktitle = {Proceedings of the 2024 {CHI} {Conference} on {Human} {Factors} in {Computing} {Systems}},
	publisher = {Association for Computing Machinery},
	author = {Liu, Yiren and Chen, Si and Cheng, Haocong and Yu, Mengxia and Ran, Xiao and Mo, Andrew and Tang, Yiliu and Huang, Yun},
	month = may,
	year = {2024},
	pages = {1--25},
	 }

@inproceedings{zhang_explaining_2024,
	title = {Explaining the {Wait}: {How} {Justifying} {Chatbot} {Response} {Delays} {Impact} {User} {Trust}},
	isbn = {979-8-4007-0511-3},
	shorttitle = {Explaining the {Wait}},
	url = {https://dl.acm.org/doi/10.1145/3640794.3665550},
	doi = {10.1145/3640794.3665550},
	language = {en},
	urldate = {2025-05-12},
	booktitle = {{ACM} {Conversational} {User} {Interfaces} 2024},
	publisher = {ACM},
	author = {Zhang, Zhengquan and Tsiakas, Konstantinos and Schneegass, Christina},
	month = jul,
	year = {2024},
	pages = {1--16},
}

@inproceedings{wang_task_2024,
	address = {New York, NY, USA},
	series = {{CHIIR} '24},
	title = {Task {Supportive} and {Personalized} {Human}-{Large} {Language} {Model} {Interaction}: {A} {User} {Study}},
	isbn = {979-8-4007-0434-5},
	shorttitle = {Task {Supportive} and {Personalized} {Human}-{Large} {Language} {Model} {Interaction}},
	url = {https://dl.acm.org/doi/10.1145/3627508.3638344},
	doi = {10.1145/3627508.3638344},
	
	urldate = {2025-05-12},
	booktitle = {Proceedings of the 2024 {Conference} on {Human} {Information} {Interaction} and {Retrieval}},
	publisher = {Association for Computing Machinery},
	author = {Wang, Ben and Liu, Jiqun and Karimnazarov, Jamshed and Thompson, Nicolas},
	month = mar,
	year = {2024},
	pages = {370--375},
	 }

@book{davenport_thinking_2005,
	title = {Thinking for a living : how to get better performance and results from knowledge workers},
	isbn = {978-1-59139-423-5},
	shorttitle = {Thinking for a living},
	language = {eng},
	urldate = {2025-06-21},
	publisher = {Harvard Business School Press},
	author = {Davenport, Thomas H.},
	year = {2005},
	}

@article{shah_using_2025,
	title = {Using {Large} {Language} {Models} to {Generate}, {Validate}, and {Apply} {User} {Intent} {Taxonomies}},
	issn = {1559-1131},
	url = {https://dl.acm.org/doi/10.1145/3732294},
	doi = {10.1145/3732294},
	urldate = {2025-06-21},
	journal = {ACM Trans. Web},
	author = {Shah, Chirag and White, Ryen and Andersen, Reid and Buscher, Georg and Counts, Scott and Das, Sarkar and Montazer, Ali and Manivannan, Sathish and Neville, Jennifer and Rangan, Nagu and Safavi, Tara and Suri, Siddharth and Wan, Mengting and Wang, Leijie and Yang, Longqi},
	month = may,
	year = {2025},
	  }

@inproceedings{gao_taxonomy_2024,
	address = {New York, NY, USA},
	series = {{CHI} {EA} '24},
	title = {A {Taxonomy} for {Human}-{LLM} {Interaction} {Modes}: {An} {Initial} {Exploration}},
	isbn = {979-8-4007-0331-7},
	shorttitle = {A {Taxonomy} for {Human}-{LLM} {Interaction} {Modes}},
	url = {https://dl.acm.org/doi/10.1145/3613905.3650786},
	doi = {10.1145/3613905.3650786},
	
	urldate = {2025-06-24},
	booktitle = {Extended {Abstracts} of the {CHI} {Conference} on {Human} {Factors} in {Computing} {Systems}},
	publisher = {Association for Computing Machinery},
	author = {Gao, Jie and Gebreegziabher, Simret Araya and Choo, Kenny Tsu Wei and Li, Toby Jia-Jun and Perrault, Simon Tangi and Malone, Thomas W},
	month = may,
	year = {2024},
	pages = {1--11},
	 }

@inproceedings{lee_while_2025,
	address = {New York, NY, USA},
	series = {{CHI} {EA} '25},
	title = {While {We} {Wait}... {How} {Users} {Perceive} {Waiting} {Times} and {Generation} {Cues} during {AI} {Image} {Generation}},
	isbn = {979-8-4007-1395-8},
	url = {https://doi.org/10.1145/3706599.3719725},
	doi = {10.1145/3706599.3719725},
	
	urldate = {2025-08-10},
	booktitle = {Proceedings of the {Extended} {Abstracts} of the {CHI} {Conference} on {Human} {Factors} in {Computing} {Systems}},
	publisher = {Association for Computing Machinery},
	author = {Lee, Hui Min and Yadav, Davis and Lee, Sangwook and Govindarazan, Keerthana and Chen, Cheng and Sundar, S. Shyam},
	month = apr,
	year = {2025},
	pages = {1--8},
}

@article{liu_effects_2014,
	title = {The {Effects} of {Interactive} {Latency} on {Exploratory} {Visual} {Analysis}},
	volume = {20},
	issn = {1941-0506},
	url = {https://ieeexplore.ieee.org/document/6876022},
	doi = {10.1109/TVCG.2014.2346452},
	
	number = {12},
	urldate = {2025-08-10},
	journal = {IEEE Transactions on Visualization and Computer Graphics},
	author = {Liu, Zhicheng and Heer, Jeffrey},
	month = dec,
	year = {2014},
	pages = {2122--2131},
	 }

@article{dillion_ai_2025,
	title = {{AI} language model rivals expert ethicist in perceived moral expertise},
	volume = {15},
	copyright = {2025 The Author(s)},
	issn = {2045-2322},
	url = {https://www.nature.com/articles/s41598-025-86510-0},
	doi = {10.1038/s41598-025-86510-0},
	
	language = {en},
	number = {1},
	urldate = {2025-08-11},
	journal = {Scientific Reports},
	author = {Dillion, Danica and Mondal, Debanjan and Tandon, Niket and Gray, Kurt},
	month = feb,
	year = {2025},
	  pages = {4084},
	 }

@article{tam_framework_2024,
	title = {A framework for human evaluation of large language models in healthcare derived from literature review},
	volume = {7},
	issn = {2398-6352},
	url = {https://www.ncbi.nlm.nih.gov/pmc/articles/PMC11437138/},
	doi = {10.1038/s41746-024-01258-7},
	
	urldate = {2025-08-11},
	journal = {NPJ Digital Medicine},
	author = {Tam, Thomas Yu Chow and Sivarajkumar, Sonish and Kapoor, Sumit and Stolyar, Alisa V. and Polanska, Katelyn and McCarthy, Karleigh R. and Osterhoudt, Hunter and Wu, Xizhi and Visweswaran, Shyam and Fu, Sunyang and Mathur, Piyush and Cacciamani, Giovanni E. and Sun, Cong and Peng, Yifan and Wang, Yanshan},
	month = sep,
	year = {2024},
	pmid = {39333376},
	pmcid = {PMC11437138},
	pages = {258},
}

@article{lee2023evaluating,
    title={Evaluating Human-Language Model Interaction},
    author={Mina Lee and Megha Srivastava and Amelia Hardy and John Thickstun and Esin Durmus and Ashwin Paranjape and Ines Gerard-Ursin and Xiang Lisa Li and Faisal Ladhak and Frieda Rong and Rose E Wang and Minae Kwon and Joon Sung Park and Hancheng Cao and Tony Lee and Rishi Bommasani and Michael S. Bernstein and Percy Liang},
    journal={Transactions on Machine Learning Research},
    issn={2835-8856},
    year={2023},
    url={https://openreview.net/forum?id=hjDYJUn9l1},
}

@misc{lenth_emmeans_2025,
	title = {emmeans: {Estimated} {Marginal} {Means}, aka {Least}-{Squares} {Means}},
	copyright = {GPL-2 {\textbar} GPL-3},
	shorttitle = {emmeans},
	url = {https://cran.r-project.org/web/packages/emmeans/index.html},
	urldate = {2025-08-12},
	author = {Lenth, Russell V. and Banfai, Balazs and Bolker, Ben and Buerkner, Paul and Giné-Vázquez, Iago and Herve, Maxime and Jung, Maarten and Love, Jonathon and Miguez, Fernando and Piaskowski, Julia and Riebl, Hannes and Singmann, Henrik},
	month = jul,
	year = {2025},
}

@inproceedings{subramonyam_bridging_2024,
	address = {New York, NY, USA},
	series = {{CHI} '24},
	title = {Bridging the {Gulf} of {Envisioning}: {Cognitive} {Challenges} in {Prompt} {Based} {Interactions} with {LLMs}},
	isbn = {979-8-4007-0330-0},
	shorttitle = {Bridging the {Gulf} of {Envisioning}},
	url = {https://dl.acm.org/doi/10.1145/3613904.3642754},
	doi = {10.1145/3613904.3642754},
	
	urldate = {2025-08-14},
	booktitle = {Proceedings of the 2024 {CHI} {Conference} on {Human} {Factors} in {Computing} {Systems}},
	publisher = {Association for Computing Machinery},
	author = {Subramonyam, Hari and Pea, Roy and Pondoc, Christopher and Agrawala, Maneesh and Seifert, Colleen},
	month = may,
	year = {2024},
	pages = {1--19},
	 }

@misc{inan_better_2025,
	title = {Better {Slow} than {Sorry}: {Introducing} {Positive} {Friction} for {Reliable} {Dialogue} {Systems}},
	shorttitle = {Better {Slow} than {Sorry}},
	url = {http://arxiv.org/abs/2501.17348},
	doi = {10.48550/arXiv.2501.17348},
	
	urldate = {2025-08-24},
	publisher = {arXiv},
	author = {İnan, Mert and Sicilia, Anthony and Dey, Suvodip and Dongre, Vardhan and Srinivasan, Tejas and Thomason, Jesse and Tür, Gökhan and Hakkani-Tür, Dilek and Alikhani, Malihe},
	month = jan,
	year = {2025},
	  }

@inproceedings{chen_exploring_2024,
	address = {Berlin, Heidelberg},
	title = {Exploring a {Behavioral} {Model} of “{Positive} {Friction}” in {Human}-{AI} {Interaction}},
	isbn = {978-3-031-61352-4},
	url = {https://doi.org/10.1007/978-3-031-61353-1_1},
	doi = {10.1007/978-3-031-61353-1_1},
	
	urldate = {2025-08-23},
	booktitle = {Design, {User} {Experience}, and {Usability}: 13th {International} {Conference}, {DUXU} 2024, {Washington}, {DC}, {USA}, {June} 29–{July} 4, 2024, {Proceedings}, {Part} {II}},
	publisher = {Springer-Verlag},
	author = {Chen, Zeya and Schmidt, Ruth},
	month = jun,
	year = {2024},
	pages = {3--22},
}

@online{stackpole_help_2024,
	title = {To help improve the accuracy of generative {AI}, add speed bumps},
	url = {https://mitsloan.mit.edu/ideas-made-to-matter/to-help-improve-accuracy-generative-ai-add-speed-bumps},
	urldate = {2025-08-24},
	organization = {MIT Sloan},
	author = {Stackpole, Beth},
	year = {2024},
	 }

@book{card_psychology_1983,
	address = {USA},
	title = {The {Psychology} of {Human}-{Computer} {Interaction}},
	isbn = {0-89859-243-7},
	publisher = {L. Erlbaum Associates Inc.},
	author = {Card, Stuart K. and Newell, Allen and Moran, Thomas P.},
	year = {1983},
}

@misc{lin_sleep-time_2025,
	title = {Sleep-time {Compute}: {Beyond} {Inference} {Scaling} at {Test}-time},
	shorttitle = {Sleep-time {Compute}},
	url = {http://arxiv.org/abs/2504.13171},
	doi = {10.48550/arXiv.2504.13171},
	urldate = {2025-08-25},
	publisher = {arXiv},
	author = {Lin, Kevin and Snell, Charlie and Wang, Yu and Packer, Charles and Wooders, Sarah and Stoica, Ion and Gonzalez, Joseph E.},
	month = apr,
	year = {2025},
	  }

@article{bucinca_trust_2021,
	title = {To {Trust} or to {Think}: {Cognitive} {Forcing} {Functions} {Can} {Reduce} {Overreliance} on {AI} in {AI}-assisted {Decision}-making},
	volume = {5},
	shorttitle = {To {Trust} or to {Think}},
	url = {https://dl.acm.org/doi/10.1145/3449287},
	doi = {10.1145/3449287},
	
	number = {CSCW1},
	urldate = {2025-08-26},
	journal = {Proc. ACM Hum.-Comput. Interact.},
	author = {Buçinca, Zana and Malaya, Maja Barbara and Gajos, Krzysztof Z.},
	month = apr,
	year = {2021},
	pages = {188:1--188:21},
	 }

@article{park_slow_2019,
	title = {A {Slow} {Algorithm} {Improves} {Users}' {Assessments} of the {Algorithm}'s {Accuracy}},
	volume = {3},
	url = {https://dl.acm.org/doi/10.1145/3359204},
	doi = {10.1145/3359204},
	
	number = {CSCW},
	urldate = {2025-08-26},
	journal = {Proc. ACM Hum.-Comput. Interact.},
	author = {Park, Joon Sung and Barber, Rick and Kirlik, Alex and Karahalios, Karrie},
	month = nov,
	year = {2019},
	pages = {102:1--102:15},
	 }

@inproceedings{liao_questioning_2020,
	title = {Questioning the {AI}: {Informing} {Design} {Practices} for {Explainable} {AI} {User} {Experiences}},
	shorttitle = {Questioning the {AI}},
	url = {http://arxiv.org/abs/2001.02478},
	doi = {10.1145/3313831.3376590},
	
	urldate = {2025-08-27},
	booktitle = {Proceedings of the 2020 {CHI} {Conference} on {Human} {Factors} in {Computing} {Systems}},
	author = {Liao, Q. Vera and Gruen, Daniel and Miller, Sarah},
	month = apr,
	year = {2020},
	  pages = {1--15},
	 }

@incollection{norman_observations_1983,
  author    = {Norman, Donald A.},
  title     = {Some observations on mental models},
  booktitle = {Mental Models},
  editor    = {Gentner, Dedre and Stevens, Albert L.},
  publisher = {Lawrence Erlbaum Associates},
  year      = {1983},
  pages     = {7--14} 
}

@misc{riemer_time_2023,
	title = {Time and {Timing} in {Human}-{Computer} {Interaction}},
	publisher = {GI},
	author = {Riemer, Martin and Bogon, Johanna and Rußwinkel, Nele and Henze, Niels and Wiese, Eva and Halbhuber, David and Thomaschke, Roland},
	year = {2023},
	doi = {10.18420/muc2023-mci-ws05-106},
	}

@article{nass_machines_2000,
	title = {Machines and {Mindlessness}: {Social} {Responses} to {Computers}},
	volume = {56},
	issn = {0022-4537, 1540-4560},
	shorttitle = {Machines and {Mindlessness}},
	url = {https://spssi.onlinelibrary.wiley.com/doi/10.1111/0022-4537.00153},
	doi = {10.1111/0022-4537.00153},
	
	language = {en},
	number = {1},
	urldate = {2025-08-27},
	journal = {Journal of Social Issues},
	author = {Nass, Clifford and Moon, Youngme},
	month = jan,
	year = {2000},
	pages = {81--103},
	 }

@inproceedings{liao_designing_2022,
	address = {New York, NY, USA},
	series = {{FAccT} '22},
	title = {Designing for {Responsible} {Trust} in {AI} {Systems}: {A} {Communication} {Perspective}},
	isbn = {978-1-4503-9352-2},
	shorttitle = {Designing for {Responsible} {Trust} in {AI} {Systems}},
	url = {https://dl.acm.org/doi/10.1145/3531146.3533182},
	doi = {10.1145/3531146.3533182},
	
	urldate = {2025-08-28},
	booktitle = {Proceedings of the 2022 {ACM} {Conference} on {Fairness}, {Accountability}, and {Transparency}},
	publisher = {Association for Computing Machinery},
	author = {Liao, Q. Vera and Sundar, S. Shyam},
	month = jun,
	year = {2022},
	pages = {1257--1268},
	 }

@article{gnewuch_chatbot_2018,
	title = {“{The} {Chatbot} is typing ...” – {The} {Role} of {Typing} {Indicators} in {Human}-{Chatbot} {Interaction}},
	url = {https://aisel.aisnet.org/sighci2018/14},
	journal = {SIGHCI 2018 Proceedings},
	author = {Gnewuch, Ulrich and Morana, Stefan and Adam, Marc and Maedche, Alexander},
	month = dec,
	year = {2018},
	 }

@article{thum_standardized_1995,
	title = {Standardized task strain and system response times in human-computer interaction},
	volume = {38},
	issn = {0014-0139},
	doi = {10.1080/00140139508925192},
	
	language = {eng},
	number = {7},
	journal = {Ergonomics},
	author = {Thum, M. and Boucsein, W. and Kuhmann, W. and Ray, W. J.},
	month = jul,
	year = {1995},
	pmid = {7635125},
	pages = {1342--1351},
}

@misc{brachman_current_2025,
	title = {Current and {Future} {Use} of {Large} {Language} {Models} for {Knowledge} {Work}},
	url = {http://arxiv.org/abs/2503.16774},
	doi = {10.48550/arXiv.2503.16774},
	
	urldate = {2025-08-28},
	publisher = {arXiv},
	author = {Brachman, Michelle and El-Ashry, Amina and Dugan, Casey and Geyer, Werner},
	month = mar,
	year = {2025},
	  }

@inproceedings{myer_towards_2002,
	address = {New York, NY, USA},
	series = {{CHI} {EA} '02},
	title = {Towards time design: pacing of hypertext navigation by system response times},
	isbn = {978-1-58113-454-4},
	shorttitle = {Towards time design},
	url = {https://dl.acm.org/doi/10.1145/506443.506616},
	doi = {10.1145/506443.506616},
	
	urldate = {2025-08-28},
	booktitle = {{CHI} '02 {Extended} {Abstracts} on {Human} {Factors} in {Computing} {Systems}},
	publisher = {Association for Computing Machinery},
	author = {Myer, Herbert A. and Hildebrandt, Michael},
	month = apr,
	year = {2002},
	pages = {824--825},
	 }

@article{kohlisch_system_1997,
	title = {System response time and readiness for task execution the optimum duration of inter-task delays},
	volume = {40},
	issn = {0014-0139},
	url = {https://doi.org/10.1080/001401397188143},
	doi = {10.1080/001401397188143},
	number = {3},
	urldate = {2025-08-28},
	journal = {Ergonomics},
	author = {Kohlisch, Olaf and Kuhmann, Werner},
	year = {1997},
	pages = {265--280},
}

@article{thomaschke_predictivity_2014,
	title = {Predictivity of system delays shortens human response time},
	volume = {72},
	issn = {1071-5819},
	url = {https://www.sciencedirect.com/science/article/pii/S1071581913001997},
	doi = {10.1016/j.ijhcs.2013.12.004},
	
	number = {3},
	urldate = {2025-08-28},
	journal = {International Journal of Human-Computer Studies},
	author = {Thomaschke, Roland and Haering, Carola},
	month = mar,
	year = {2014},
	pages = {358--365},
	 }

@article{hwang_when_2019,
	title = {When {Delayed} in a {Hurry}: {Interpretations} of {Response} {Delays} in {Time}-{Sensitive} {Instant} {Messaging}},
	volume = {3},
	shorttitle = {When {Delayed} in a {Hurry}},
	url = {https://dl.acm.org/doi/10.1145/3361115},
	doi = {10.1145/3361115},
	
	number = {GROUP},
	urldate = {2025-08-29},
	journal = {Proc. ACM Hum.-Comput. Interact.},
	author = {Hwang, Sun Young and Khojasteh, Negar and Fussell, Susan R.},
	month = dec,
	year = {2019},
	pages = {234:1--234:20},
	 }

@article{kruger_effort_2004,
	title = {The effort heuristic},
	volume = {40},
	issn = {0022-1031},
	url = {https://www.sciencedirect.com/science/article/pii/S0022103103000659},
	doi = {10.1016/S0022-1031(03)00065-9},
	
	number = {1},
	urldate = {2025-08-29},
	journal = {Journal of Experimental Social Psychology},
	author = {Kruger, Justin and Wirtz, Derrick and Van Boven, Leaf and Altermatt, T. William},
	month = jan,
	year = {2004},
	pages = {91--98},
	 }

@article{logg_algorithm_2019,
	title = {Algorithm appreciation: {People} prefer algorithmic to human judgment},
	volume = {151},
	issn = {0749-5978},
	shorttitle = {Algorithm appreciation},
	url = {https://www.sciencedirect.com/science/article/pii/S0749597818303388},
	doi = {10.1016/j.obhdp.2018.12.005},
	
	urldate = {2025-08-29},
	journal = {Organizational Behavior and Human Decision Processes},
	author = {Logg, Jennifer M. and Minson, Julia A. and Moore, Don A.},
	month = mar,
	year = {2019},
	pages = {90--103},
}

@inproceedings{riedl_system_2018,
	address = {Berlin, Heidelberg},
	title = {System {Response} {Time} as a {Stressor} in a {Digital} {World}: {Literature} {Review} and {Theoretical} {Model}},
	isbn = {978-3-319-91715-3},
	shorttitle = {System {Response} {Time} as a {Stressor} in a {Digital} {World}},
	url = {https://doi.org/10.1007/978-3-319-91716-0_14},
	doi = {10.1007/978-3-319-91716-0_14},
	
	urldate = {2025-08-28},
	booktitle = {{HCI} in {Business}, {Government}, and {Organizations}: 5th {International} {Conference}, {HCIBGO} 2018, {Las} {Vegas}, {NV}, {USA}, {July} 15-20, 2018, {Proceedings}},
	publisher = {Springer-Verlag},
	author = {Riedl, René and Fischer, Thomas},
	month = jul,
	year = {2018},
	pages = {175--186},
}

@article{dabrowski_40_2011,
  title = {40 years of searching for the best computer system response time},
  author = {Dabrowski, Jim and Munson, Ethan V.},
  journal = {Interacting with Computers},
  volume = {23},
  number = {5},
  pages = {555--564},
  year = {2011},
  doi = {10.1016/j.intcom.2011.05.008},
  issn = {1873-7951}
}

@article{mendel2025laypeople,
  title={Laypeople’s use of and attitudes toward large language models and search engines for health queries: survey study},
  author={Mendel, Tamir and Singh, Nina and Mann, Devin M and Wiesenfeld, Batia and Nov, Oded},
  journal={Journal of medical Internet research},
  volume={27},
  pages={e64290},
  year={2025},
  publisher={JMIR Publications Toronto, Canada},
    doi = {10.2196/64290}
}

@inproceedings{pu2025ideasynth,
  title={Ideasynth: Iterative research idea development through evolving and composing idea facets with literature-grounded feedback},
  author={Pu, Kevin and Feng, KJ Kevin and Grossman, Tovi and Hope, Tom and Dalvi Mishra, Bhavana and Latzke, Matt and Bragg, Jonathan and Chang, Joseph Chee and Siangliulue, Pao},
  pages={1--31},
  year={2025},
    publisher = {Association for Computing Machinery},
    address = {New York, NY, USA},
    url = {https://doi.org/10.1145/3706598.3714057},
    series = {CHI '25}
    }

@inproceedings{zhang2025friction,
  title={Friction: Deciphering Writing Feedback into Writing Revisions through LLM-Assisted Reflection},
  author={Zhang, Chao and Ju, Kexin and Bidoshi, Peter and Yen, Yu-Chun Grace and Rzeszotarski, Jeffrey M},
    year = {2025},
    pages={1--27},
    isbn = {9798400713941},
    publisher = {Association for Computing Machinery},
    address = {New York, NY, USA},
    url = {https://doi.org/10.1145/3706598.3714316},
    doi = {10.1145/3706598.3714316},
    booktitle = {Proceedings of the 2025 CHI Conference on Human Factors in Computing Systems},
    series = {CHI '25}
    }

@article{mandal2025utilization,
  title={Utilization of generative AI-drafted responses for managing patient-provider communication},
  author={Mandal, Soumik and Wiesenfeld, Batia M and Szerencsy, Adam C and Small, William R and Major, Vincent and Richardson, Safiya and Schoenthaler, Antoinette and Mann, Devin and Nov, Oded},
  journal={npj Digital Medicine},
  volume={8},
  number={1},
  pages={591},
  year={2025},
  publisher={Nature Publishing Group UK London},
  doi = {10.1038/s41746-025-01972-w},
  url = {https://doi.org/10.1038/s41746-025-01972-w}
}

@incollection{hart_development_1988,
	series = {Human {Mental} {Workload}},
	title = {Development of {NASA}-{TLX} ({Task} {Load} {Index}): {Results} of {Empirical} and {Theoretical} {Research}},
	volume = {52},
	shorttitle = {Development of {NASA}-{TLX} ({Task} {Load} {Index})},
	url = {https://www.sciencedirect.com/science/article/pii/S0166411508623869},
	urldate = {2025-08-30},
	booktitle = {Advances in {Psychology}},
	publisher = {North-Holland},
	author = {Hart, Sandra and Staveland, Lowell},
	editor = {Hancock, Peter A. and Meshkati, Najmedin},
	month = jan,
	year = {1988},
	doi = {10.1016/S0166-4115(08)62386-9},
	pages = {139--183},
	 }

@inproceedings{wobbrock_aligned_2011,
	address = {New York, NY, USA},
	series = {{CHI} '11},
	title = {The aligned rank transform for nonparametric factorial analyses using only anova procedures},
	isbn = {978-1-4503-0228-9},
	url = {https://doi.org/10.1145/1978942.1978963},
	doi = {10.1145/1978942.1978963},
	
	urldate = {2025-08-30},
	booktitle = {Proceedings of the {SIGCHI} {Conference} on {Human} {Factors} in {Computing} {Systems}},
	publisher = {Association for Computing Machinery},
	author = {Wobbrock, Jacob O. and Findlater, Leah and Gergle, Darren and Higgins, James J.},
	month = may,
	year = {2011},
	pages = {143--146},
}

@incollection{maister_psychology_1985,
    author    = {Maister, David H.},
	title = {The {Psychology} of {Waiting} {Lines}},
      booktitle = {The Service Encounter},
      editor    = {Czepiel, John A. and Solomon, Michael R. and Surprenant, Carol},
      publisher = {Lexington Books},
      address   = {Lexington, MA},
      year      = {1985},
      pages     = {113--123}
	 }

@misc{sui_stop_2025,
	title = {Stop {Overthinking}: {A} {Survey} on {Efficient} {Reasoning} for {Large} {Language} {Models}},
	shorttitle = {Stop {Overthinking}},
	url = {http://arxiv.org/abs/2503.16419},
	doi = {10.48550/arXiv.2503.16419},
	
	urldate = {2025-09-04},
	publisher = {arXiv},
	author = {Sui, Yang and Chuang, Yu-Neng and Wang, Guanchu and Zhang, Jiamu and Zhang, Tianyi and Yuan, Jiayi and Liu, Hongyi and Wen, Andrew and Zhong, Shaochen and Zou, Na and Chen, Hanjie and Hu, Xia},
	month = aug,
	year = {2025},
	  }

@article{sacks_simplest_1974,
	title = {A {Simplest} {Systematics} for the {Organization} of {Turn}-{Taking} for {Conversation}},
	volume = {50},
	issn = {0097-8507},
	url = {https://www.jstor.org/stable/412243},
	doi = {10.2307/412243},
	
	number = {4},
	urldate = {2025-09-04},
	journal = {Language},
	author = {Sacks, Harvey and Schegloff, Emanuel A. and Jefferson, Gail},
	year = {1974},
	  pages = {696--735},
	 }

@inproceedings{kuang_enhancing_2024,
	address = {New York, NY, USA},
	series = {{CHI} '24},
	title = {Enhancing {UX} {Evaluation} {Through} {Collaboration} with {Conversational} {AI} {Assistants}: {Effects} of {Proactive} {Dialogue} and {Timing}},
	isbn = {979-8-4007-0330-0},
	shorttitle = {Enhancing {UX} {Evaluation} {Through} {Collaboration} with {Conversational} {AI} {Assistants}},
	url = {https://dl.acm.org/doi/10.1145/3613904.3642168},
	doi = {10.1145/3613904.3642168},
	
	urldate = {2025-09-04},
	booktitle = {Proceedings of the 2024 {CHI} {Conference} on {Human} {Factors} in {Computing} {Systems}},
	publisher = {Association for Computing Machinery},
	author = {Kuang, Emily and Li, Minghao and Fan, Mingming and Shinohara, Kristen},
	month = may,
	year = {2024},
	pages = {1--16},
	 }

@inproceedings{yeh_how_2022,
	address = {New York, NY, USA},
	series = {{CHI} '22},
	title = {How to {Guide} {Task}-oriented {Chatbot} {Users}, and {When}: {A} {Mixed}-methods {Study} of {Combinations} of {Chatbot} {Guidance} {Types} and {Timings}},
	isbn = {978-1-4503-9157-3},
	shorttitle = {How to {Guide} {Task}-oriented {Chatbot} {Users}, and {When}},
	url = {https://dl.acm.org/doi/10.1145/3491102.3501941},
	doi = {10.1145/3491102.3501941},
	
	urldate = {2025-09-05},
	booktitle = {Proceedings of the 2022 {CHI} {Conference} on {Human} {Factors} in {Computing} {Systems}},
	publisher = {Association for Computing Machinery},
	author = {Yeh, Su-Fang and Wu, Meng-Hsin and Chen, Tze-Yu and Lin, Yen-Chun and Chang, XiJing and Chiang, You-Hsuan and Chang, Yung-Ju},
	month = apr,
	year = {2022},
	pages = {1--16},
	 }

@inproceedings{dang_choice_2023,
	address = {New York, NY, USA},
	series = {{CHI} '23},
	title = {Choice {Over} {Control}: {How} {Users} {Write} with {Large} {Language} {Models} using {Diegetic} and {Non}-{Diegetic} {Prompting}},
	isbn = {978-1-4503-9421-5},
	shorttitle = {Choice {Over} {Control}},
	url = {https://doi.org/10.1145/3544548.3580969},
	doi = {10.1145/3544548.3580969},
	
	urldate = {2025-09-07},
	booktitle = {Proceedings of the 2023 {CHI} {Conference} on {Human} {Factors} in {Computing} {Systems}},
	publisher = {Association for Computing Machinery},
	author = {Dang, Hai and Goller, Sven and Lehmann, Florian and Buschek, Daniel},
	month = apr,
	year = {2023},
	pages = {1--17},
	 }

@article{haigh_role_2021,
    title = {The role of {Weber}’s law in human time perception},
    volume = {83},
    issn = {1943-393X},
    url = {https://doi.org/10.3758/s13414-020-02128-6},
    doi = {10.3758/s13414-020-02128-6},
    
    language = {en},
    number = {1},
    urldate = {2024-10-07},
    journal = {Attention, Perception, \& Psychophysics},
    author = {Haigh, Andrew and Apthorp, Deborah and Bizo, Lewis A.},
    month = jan,
    year = {2021},
    pages = {435--447},
}

@article{araujo_living_2018,
    title = {Living up to the chatbot hype: {The} influence of anthropomorphic design cues and communicative agency framing on conversational agent and company perceptions},
    volume = {85},
    issn = {0747-5632},
    shorttitle = {Living up to the chatbot hype},
    url = {https://www.sciencedirect.com/science/article/pii/S0747563218301560},
    doi = {10.1016/j.chb.2018.03.051},
    
    urldate = {2025-09-11},
    journal = {Computers in Human Behavior},
    author = {Araujo, Theo},
    month = aug,
    year = {2018},
    pages = {183--189},
}

\appendix

\section{Experimental Tasks}
\label{appendix-Experimental tasks}

    \subsection{Creation Tasks}

    \paragraph{\textbf{Task 1: Slogan Generation + Rewrite}}  
    Imagine you are creating a promotional slogan for a new online platform called \emph{PeerLink}, designed to help high school students collaborate on group projects in real time using shared documents, chat, and task tracking.  
    Your slogan should be aimed at teachers and school leaders who might consider recommending this tool to their students.  
    Use the AI assistant to brainstorm, refine, or evaluate your slogan ideas.  
    \emph{In the response field, submit your finalized slogan, along with an explanation of why you chose it.}  

    \paragraph{\textbf{Task 2: Scenario-Based Brainstorming}}  
    A group of high school students are working together in after-school study sessions, but they often get distracted or lose focus while trying to prepare for exams and complete assignments.  
    Use the AI assistant to help you brainstorm and refine activity ideas that could assist the students with staying on task.  
    \emph{In the response field, describe one idea you have developed and explain how you would adapt it for this group.}
    
    \paragraph{\textbf{Task 3: Gap-Filling Writing}}  
    Below is an excerpt of an article titled \emph{``Why Is Learning a Second Language Important?''} The introduction and conclusion are already written, but the middle section is missing. This missing section should provide one or two key reasons someone might benefit from learning a new language.  
    Use the AI assistant to help you brainstorm and refine your ideas.  
    \emph{In the response field, write out 3-4 sentences to complete the article's middle section.}

\subsection{Advice Tasks}
    
    \paragraph{\textbf{Task 1: Personalized Recommendation}}  
    A coworker shares that they’ve been struggling to stay focused while working from home. They mention that distractions like social media, house chores, or lack of routine are getting in the way of their productivity.  
    Use the AI assistant to suggest practical strategies they can try to improve their focus.  
    \emph{In the response field, describe one chosen strategy and explain why you think this strategy is effective and realistic for someone working remotely.}  
    
    \paragraph{\textbf{Task 2: Decision Memo Review}}  
    Someone has drafted a short decision memo comparing two career options: (1) accepting a promotion at their current company, and (2) taking a job offer from another company with higher pay and more flexibility.  
    Use the AI assistant to help assess whether the memo considers all relevant factors that people might weigh in a career decision.  
    \emph{In the response field, list any additional factors or information that you think are important to consider when making a career decision like this. Explain how including these considerations might improve the memo and help make the decision process clearer.}  
    
    \paragraph{\textbf{Task 3: Email Reply Revision}}  
    A friend has emailed you asking for advice:  
    \emph{``I’m thinking about quitting my job in marketing to become a full-time artist. I’ve been doing freelance illustration on the side for a few years, and it’s what I really love. But I’m nervous about the financial uncertainty. What do you think I should do?''}  
    
    You’ve drafted a quick reply:  
    \emph{``I think you should go for it. You clearly love art and life is short. Worst case, you can always go back to marketing.''} Use the AI assistant to help you improve this reply. Make it more thoughtful, balanced, and emotionally supportive.  
    \emph{In the response field, submit your revised version of the email reply.}

\onecolumn

\section{Participant demographics}
\label{participant demographics}
    \begin{table}[htbp]
    \centering
    \caption{Participant demographics and AI usage.}
    \begingroup
    \renewcommand{\arraystretch}{0.9}
    \Description{Three-column table summarizing participant demographics and primary AI-assistant usage (Dimension, Category, Participants). Gender: Female 126 (52.5\%), Male 112 (46.7\%), Non-binary/third gender 2 (0.8\%). Age: 18–24 = 17 (7.1\%), 25–34 = 63 (26.2\%), 35–44 = 77 (32.1\%; largest), 45–54 = 48 (20.0\%), 55+ = 35 (14.6\%). Education: High school 62 (26.7\%), Associate/Technical 18 (7.5\%), Bachelor’s 99 (41.2\%; largest), Advanced degree 61 (25.4\%). Top occupational industries: Computer and Mathematical 40 (16.7\%), Business and Financial Operations 36 (15.0\%), Educational Instruction and Library 24 (10.0\%), Arts/Design/Entertainment/Sports/Media 17 (7.1\%), Management 16 (6.7\%). Primary AI assistant (top 5): ChatGPT (OpenAI) 187 (77.9\%; dominant), Google Gemini 30 (12.5\%), Claude.ai 6 (2.5\%), Perplexity.ai 6 (2.5\%), Grok 4 (1.7\%).}
    \begin{tabular}{lll}
    \hline
    \textbf{Dimension} & \textbf{Category} & \textbf{Participants} \\
    \hline
    Gender Identity & Female & 126 (52.5\%) \\
                    & Male & 112 (46.7\%) \\
                    & Non-binary / third gender & 2 (0.8\%) \\
                   
    \hline
    Age Range       & 18--24 & 17 (7.1\%) \\
                    & 25--34 & 63 (26.2\%) \\
                    & 35--44 & 77 (32.1\%) \\
                    & 45--54 & 48 (20.0\%) \\
                    & 55+ & 35 (14.6\%) \\
    \hline
    Education       & High school & 62 (26.7\%) \\
                    
                    & Associate/Technical degree & 18 (7.5\%) \\
                    & Bachelor’s degree & 99 (41.2\%) \\
                    & Advanced degree & 61 (25.4\%) \\
   
    \hline
    Occupational Industry (top 5) & Computer and Mathematical & 40 (16.7\%) \\
                    & Business and Financial Operations & 36 (15.0\%) \\
                    & Educational Instruction and Library & 24 (10.0\%) \\
                    & Arts, Design, Entertainment, Sports, and Media & 17 (7.1\%) \\
                    & Management & 16 (6.7\%) \\
     \hline
        Primary AI assistant (top 5)* & ChatGPT (OpenAI) & 187 (77.9\%) \\
                        & Google Gemini & 30 (12.5\%) \\
                        & Claude.ai & 6 (2.5\%) \\
                        & Perplexity.ai & 6 (2.5\%) \\
                        & Grok & 4 (1.7\%) \\
    \hline
    \end{tabular}
    \par\endgroup
    \vspace{2mm}
    {\footnotesize \textit{\textbf{*} Participants were asked: ``Which AI assistant tool do you most regularly use for knowledge tasks?''}\par}
    \end{table}

\section{Qualitative themes}
\label{appendix-Qualitative themes}

\begin{table*}[htbp]
\centering
\caption{Seven themes from participants’ open-ended responses. Counts reflect responses coded with each theme; multiple codes per response were allowed, so percentages do not sum to 100\%.}

\Description{Three-column table titled “Seven themes from participants’ open-ended responses.” Columns: Theme (with count and percent), Definition, and Example Quote. Themes and counts: No/Minimal Impact (n=63, 45.0\%); Inferred AI Deliberation (n=43, 30.7\%); Quality-Contingent Value (n=19, 13.6\%); Self-Regulatory Use (n=5, 3.6\%); Negative Affective Cost (n=10, 7.1\%); System Reliability Concern (n=20, 14.3\%); Interaction Strategy Adaptation (n=12, 8.6\%). Definitions summarize how delay was interpreted: no substantive impact; seen as AI “thinking”; acceptable if quality improves; time used to plan; frustrating or distracting; raised technical doubts; or prompted strategy changes. Each row includes an illustrative participant quote.}

\label{tab:themes}
\small
\begin{tabular}{p{4cm} p{6.5cm} p{6cm}}
\hline
\textbf{Theme (Count)} & \textbf{Definition} & \textbf{Example Quote} \\
\hline
\makecell[l]{\\ No/Minimal Impact \\ \textit{($n=63$, 45.0\%)}} 
& Delay is noticed but has no substantive effect on process or quality; may involve minor irritation or inconvenience. 
& ``Did not influence the way I interacted...It was just minimally annoying.'' \\
    
\makecell[l]{\\ Inferred AI Deliberation \\ \textit{($n=43$, 30.7\%)}} 
& Delay is attributed to AI cognition. As a result, participants may expect higher-quality or more thoughtful responses. 
& ``It was gathering information to formulate a thoughtful response...made me think that I would get a well planned and well written response.'' \\
    
\makecell[l]{\\ Quality-Contingent Value \\ \textit{($n=19$, 13.6\%)}} 
& Delay is acceptable when the response is clear, detailed, or accurate; expectations of output value scale with delay. 
& ``If the AI ultimately delivered a better response more detailed, accurate, or insightful than expected I’d likely view the wait as worthwhile.'' \\
    
\makecell[l]{\\ Self-Regulatory Use \\ \textit{($n=5$, 3.6\%)}} 
& Delay is framed as useful time for reflection, planning, or preparing the next query. 
& ``I had longer to contemplate what my next query would be.'' \\
    
\makecell[l]{\\ Negative Affective Cost \\ \textit{($n=10$, 7.1\%)}} 
& Delay produces negative emotional consequences such as impatience, frustration, distraction, or loss of focus. 
& ``It made me slightly anxious and distracted me from the questions...I may have missed one of the details.'' \\
    
\makecell[l]{\\ System Reliability Concern \\ \textit{($n=20$, 14.3\%)}} 
& Delay raises technical doubts about system/LLM reliability or efficiency. 
& ``The response time made me question if my internet cut off.'' \\
    
\makecell[l]{\\ Interaction Strategy Adaptation \\ \textit{($n=12$, 8.6\%)}} 
& Delay leads participants to change their interaction strategies, e.g., fewer prompts or front-loading information. 
& ``I tried to minimize the number of prompts per task...'' \\
\hline
\end{tabular}
\end{table*}

\end{document}